\documentclass{jfm}
\usepackage{graphicx}
\usepackage{epstopdf, epsfig}
\usepackage{mathtools}
\usepackage{amsmath}
\usepackage{todonotes}
\usepackage{appendix}
\usepackage{cleveref}
\usepackage{float}
\usepackage{soul}
\usepackage{tabularx}
\definecolor{green}{RGB}{34,150,34}
\usepackage{soul}
\usepackage{gensymb}
\usepackage[labelformat=simple]{subcaption}

\shorttitle{A new Lagrangian drift mechanism due to current--bathymetry interactions}
\shortauthor{A. Gupta \& A. Guha}

\title{A new Lagrangian drift mechanism due to current--bathymetry interactions: applications in coastal cross-shelf transport}

\author{Akanksha Gupta\aff{1,2} \corresp{\email{akanksha.gupta.iitbhu@gmail.com}} \and Anirban Guha\aff{2}}

\affiliation{\aff{1}Department of Mechanical Engineering,
Indian Institute of Technology, Kanpur, U.P. 208016, India.\\
\aff{2} School of Science and Engineering, University of Dundee, DD1 4HN, U.K.}

\def\ee{{\rm e}}

\newcommand{\vect}[1]{\boldsymbol{#1}}

\begin{document}

\maketitle

\begin{abstract}
We show that in free surface flows, a uniform, streamwise current over small-amplitude   wavy bottom topography generates cross-stream drift velocity. This  drift mechanism, referred to as the current-bathymetry interaction induced drift (CBIID), is specifically understood in the context of a simplified nearshore environment consisting of a uniform alongshore current, onshore propagating surface waves, and monochromatic wavy bottom making an oblique angle with the shoreline. CBIID is found to originate from the steady, non-homogeneous solution of the governing system of equations. Similar to Stokes drift induced by surface waves, CBIID also generates a compensating Eulerian return flow to satisfy the no-flux lateral boundaries, e.g.\,the shoreline. CBIID increases with an increase in particle's initial depth, bottom undulation's amplitude, and the strength of the alongshore current. Additionally,  CBIID near the free (bottom) surface increases (decreases) with an increase in bottom undulation's wavelength. Maximum CBIID is obtained for long wavelength bottom topography that approximately makes $\pi/4$ angle with the shoreline.  Unlike Stokes drift, particle excursions due to current-bathymetry interactions might not be small, hence analytical expressions based on the small-excursion approximation could be inaccurate. We provide an alternative $z$-bounded approximation, which leads to highly accurate expressions for drift velocity and time period of particles especially located near the free surface. Realistic  parametric analysis reveals that in some nearshore environments, CBIID's contribution to the net Lagrangian drift can be as important as Stokes drift, implying that CBIID can have  major implications in  cross-shelf tracer transport.
\end{abstract}



\vspace{-15mm}

{\section{Introduction}}
\label{sec:intro}
 
\begin{figure}
  \centering
  \hspace{8mm}
  \begin{subfigure}{0.35\textwidth}
  \centering
  \includegraphics[width=\linewidth]{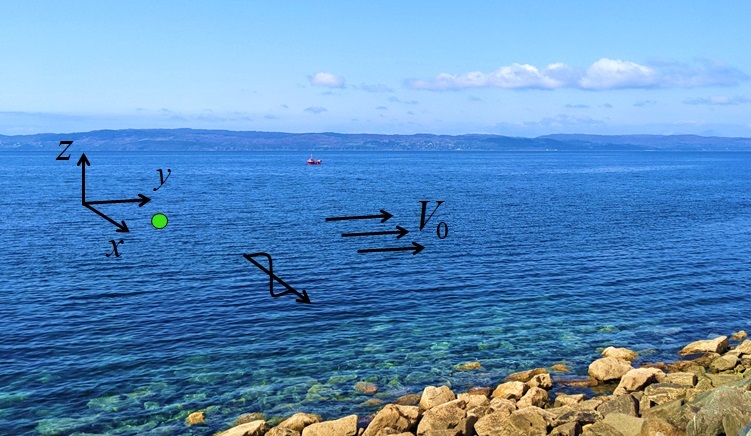}
  \vspace{-4mm}
  \caption{} \label{fig:real_schematic}
  \end{subfigure}
  \hspace{5mm}
  \centering
  \begin{subfigure}{0.53\textwidth}
  \centering
  \includegraphics[width=\linewidth]{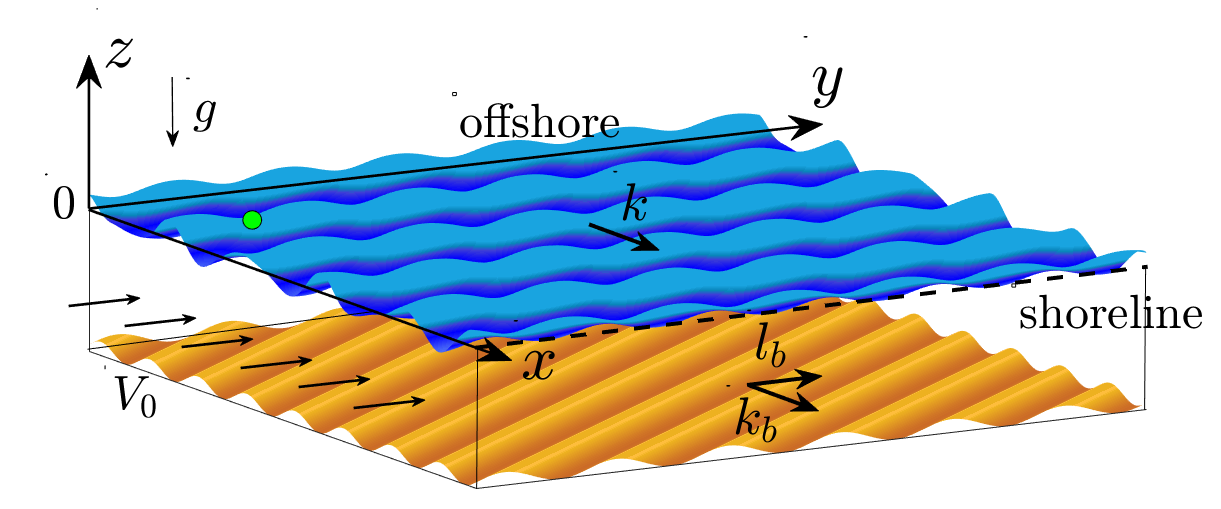}
  \vspace{-4mm}
  \caption{} \label{fig:schematic_combined}
  \end{subfigure}
  \begin{subfigure}{0.48\textwidth}
  \centering
  \includegraphics[width=\linewidth]{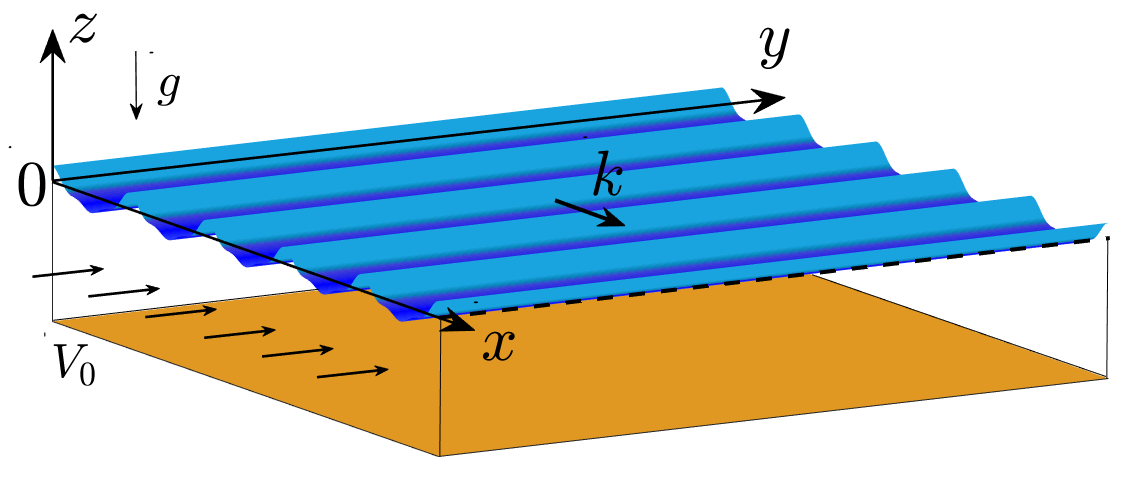}
  \caption{}  \label{fig:schematic_homogeneous}
  \end{subfigure}
  \begin{subfigure}{0.48\textwidth}
  \centering
  \includegraphics[width=\linewidth]{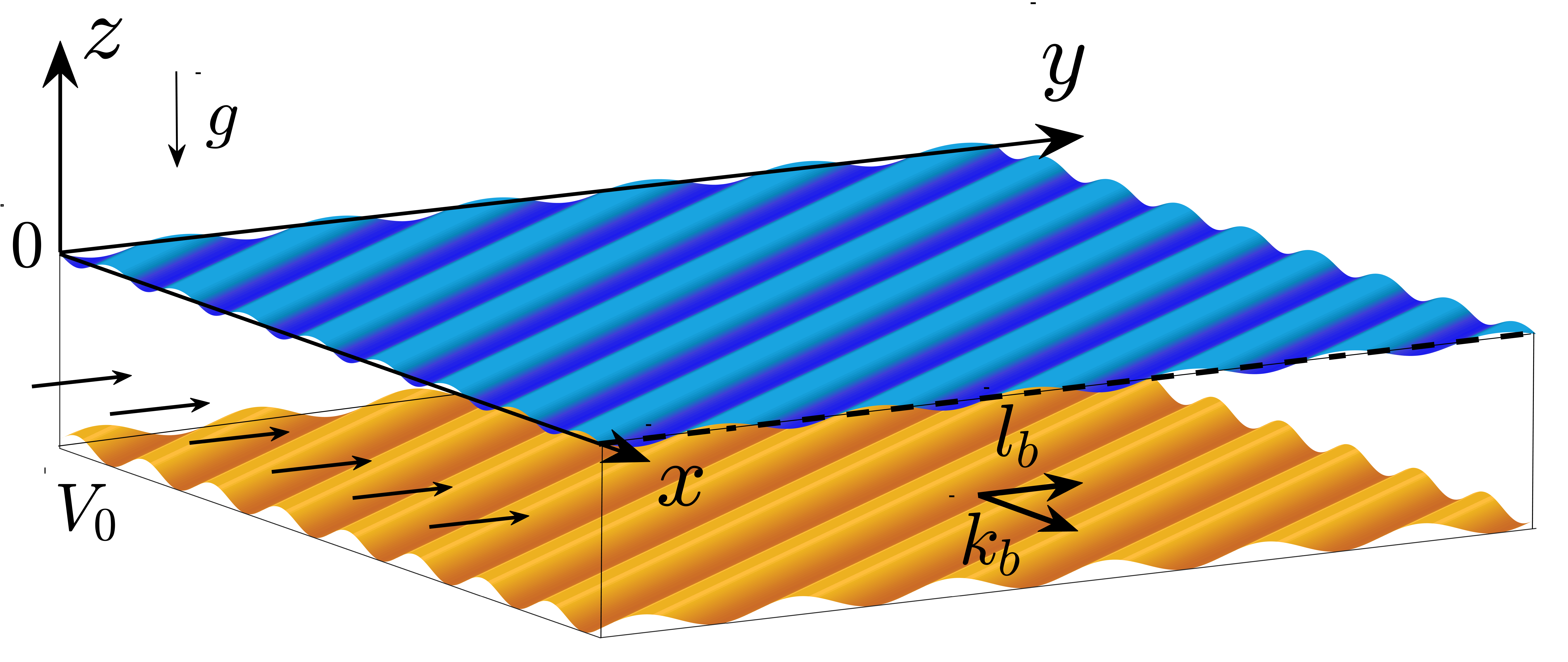}
  \vspace{-4mm}  
  \caption{} \label{fig:schematic_PI} 
  \end{subfigure}
  \caption{\footnotesize (a) A typical nearshore environment (location: Mallaig, Highlands of Scotland), and (b) the corresponding schematic diagram showing surface waves (wavenumber $K(k,0)$), alongshore current ($V_0$), wavy bottom topography (wavenumber $K_b(k_b,l_b$)) and the free surface imprint resulting from current--bathymetry interactions. Two subsets of the above situation are considered:  (c) surface waves, flat bottom and alongshore current, and (d) wavy bottom topography and alongshore current, but no surface waves. In the last case, the undulations at the free surface represent the surface imprint of the wavy seabed. The black dashed line is used for the shoreline, and the green dot in (a, b) represents a particle at the ocean surface.}  
  \label{fig:Schmatic_diagram}
\end{figure}

Rivers, estuaries, and coastal oceans are some examples of free surface flow environments that often exhibit shallow depths (few meters), moderately strong currents (up to few meters per second), and fairly complex bottom topography. The free surface flow that we will particularly focus on is the nearshore environment -- the transition region between the shoreline and the open ocean; see figure \ref{fig:Schmatic_diagram}\,({\it a\/}). The nearshore region horizontally stretches for approximately 1\,km, and consists of the surf zone (region of wave breaking) and the inner shelf (depths varying between a few meters to tens of meters) \citep{lentz2012wind,kumar2017effect}. Cross-shelf transport, i.e. the exchange of sediments, pollutants, nutrients, larvae, and pathogens between the coastal waters and the open ocean, is arguably the central problem in coastal physical oceanography \citep{lentz2008observations,brink2016cross}.  Cross-shelf currents are much weaker than  alongshore currents.  In the surf-zone, cross-shelf and alongshore currents can respectively reach up to $0.2$ m/s and  $1.5$ m/s \citep{bakhtyar2016impacts}. In the inner shelf region, cross-shelf current can typically have a magnitude of $0.01$--$0.1$ m/s, whereas the corresponding alongshore current is $0.1$--$0.5$ m/s \citep{rao2004flow}.
However, cross-shelf gradients of most properties are usually far greater than those in the alongshore direction, which causes cross-shelf exchange to dominate the rates and pathways of tracer delivery and removal on the continental shelf \citep{brink2016cross}. Onshore propagating surface waves lead to cross-shelf transport in the onshore direction via the Stokes drift mechanism -- net mass transport in the direction of surface wave propagation \citep{stokes1847}, while Eulerian return flow and  transient rip currents are some of the important mechanisms  causing offshore transport   \citep{lentz2008observations,brown2015field,kumar2017effect,o2021simulations}.

{The schematic in figure \ref{fig:Schmatic_diagram}\,({\it b\/}) provides an idealized representation of figure \ref{fig:Schmatic_diagram}\,({\it a\/}).  This idealized scenario consists of three key elements: (i) a steady, uniform, alongshore current, (ii) onshore propagating monochromatic surface waves, and (iii) a small amplitude, monochromatic bottom topography with wave-vector making an oblique angle with the shoreline. { Nearshore oblique sandbars have been observed in various locations, e.g.\, Trabucador beach, Duck beach, several Oregon beaches, St. James Island, Durras Beach, etc. \citep{ribas2003nearshore}}. According to our current understanding, the motion of a tracer parcel in the simplified set-up given in figure \ref{fig:Schmatic_diagram}\,({\it b\/}) is expected to result from two different mechanisms: Stokes drift and longshore drift (advection by the alongshore current). Hence we expect a tracer parcel  to move in a resultant direction whose streamwise component is along $+y$ (due to the longshore drift) and cross-stream component is along $+x$ (or onshore, due to the Stokes drift).
The question we ask is -- if we replace the set-up in figure \ref{fig:Schmatic_diagram}\,({\it b\/}) with a flat bathymetry (see figure \ref{fig:Schmatic_diagram}\,{\it c\/}), does it alter the trajectory of a given tracer parcel?  
The  primary objective of this paper is to show that
small amplitude   {wavy} bottom topography \emph{indeed} affects tracer trajectories, and in fact, can play a crucial role in cross-shelf { (in a generic open-channel flow, this would imply cross-stream)} tracer transport.}

{The fact that small amplitude bottom topography can impact cross-shelf tracer transport is {non-obvious}. If we assume surface waves in  figure \ref{fig:Schmatic_diagram}\,({\it a\/}) or \ref{fig:Schmatic_diagram}\,({\it b\/}) to be absent, basic fluid mechanics tells us that the tracer parcel marked by green dot will be simply advected  in the $+y$ direction by the alongshore current (i.e. undergo longshore drift).
In the presence of small amplitude  topography, we will show that an additional mechanism is at play, which can lead to cross-shelf (along $+x$ or $-x$) tracer transport. The proposed mechanism owes its existence {to the stationary waves generated due to a uniform flow over a  sinusoidal  bottom  topography \citep{thomson1886lx,lamb1932}. }
These stationary waves (or steady surface imprints of the sinusoidal wavy bottom) are shown in figure \ref{fig:Schmatic_diagram}\,({\it d\/}); they also exist in  figure \ref{fig:Schmatic_diagram}\,({\it b\/}), and can be unravelled by  removing the propagating surface waves entirely. The amplitude of these surface imprints may not be insignificant in fluvial and coastal environments owing to their shallow depths and high velocity scales, and hence can lead to non-trivial kinematics.}

The outline of the paper is as follows. In \S \ref{sec:formulation}, we provide the general mathematical formulation of the problem.  In \S \ref{sec:Case1}, we concentrate on figure \ref{fig:Schmatic_diagram}\,({\it c\/}) and onshore tracer transport of floating particles due to Stokes drift. \S \ref{sec:Case2} focuses on figure \ref{fig:Schmatic_diagram}\,({\it d\/}) and reveals a new  drift mechanism resulting from the alongshore current and  {wavy} seabed  interactions. How this drift mechanism can contribute to the cross-shelf transport, and hence affect the fate of tracer parcels, is discussed in detail. In  \S \ref{sec:case3} we discuss the set-up in  figure \ref{fig:Schmatic_diagram}\,({\it b\/}), i.e. the combined effect of surface waves,  {wavy} seabed, and alongshore current. {\S \ref{sec:casestudy} provides a comparison of the Lagrangian transport due to Stokes drift, the new drift mechanism, and their combination  for realistic parameters.} The paper is summarized and concluded in \S \ref{sec:conclusion}.


\vspace{2mm}

{\section{Mathematical formulation} \label{sec:formulation}}

We consider the three-dimensional (3D) problem of surface wave propagation over an undulating seabed in the presence of uniform background current; see figure\, \ref{fig:Schmatic_diagram}\,({\it b\/}). We assume the fluid to be  irrotational, incompressible, inviscid, and homogeneous; the domain has infinite horizontal extent but has a finite mean depth $H$. Surface tension and Coriolis effects are neglected. 
The water surface is denoted by $z=\eta(x,y,t)$; $x$ and $y$ respectively denote the cross-shelf (onshore and offshore are used respectively for positive and negative $x$-directions) and the alongshore ({i.e. streamwise}) directions, while the $z$-axis is directed upwards. The  {wavy} seabed is denoted by $z=-H+\eta_b(x,y)$, { where $|\eta_b/H| \ll 1$}.  We also consider a uniform cross-shelf current, $U_0$, and a uniform alongshore current, $V_0$.
{The fluid motion  is defined by a velocity potential, which is a combination of the velocity potential due to the uniform currents, and the perturbed velocity potential ($\phi$).}  The perturbed velocity potential satisfies the governing Laplace equation (GLE)
\begin{equation}
[\mathrm {GLE}]: \quad \phi_{,xx}+\phi_{,yy}+\phi_{,zz}=0,  \qquad {-H+\eta_b\!<\!z\!<\!\eta}  \label{eq: GE}
\end{equation}
where the comma subscript denotes partial derivative ($\phi_{,x}=\partial \phi/\partial x $). Hereafter, unless specifically mentioned, velocity potential will \emph{always} imply  perturbed quantity. 
Impenetrability condition {(ImC)} holds at the  {wavy} seabed, $z=-H+\eta_b(x,y)$,
\begin{equation}
 [\mathrm {ImC}]: \quad    \phi_{,z}-\phi_{,x} \eta_{b,x}-\phi_{,y} \eta_{b,y}=U_0 \eta_{b,x}+V_0 \eta_{b,y}. \label{eq:ImC}
\end{equation}
\noindent  ImC is a non-homogeneous equation in general, and would lead to a homogeneous solution \emph{only} when the level sets of $\eta_b$ are parallel to the background current field. 
The kinematic {(KBC)} and dynamic boundary conditions {(DBC)} at the free water surface, $z=\eta(x,y,t)$, are respectively given as 
\begin{subequations}
\begin{align}
[\mathrm {KBC}]:  \quad & \eta_{,t}+(\phi_{,x}+U_0)\eta_{,x}+(\phi_{,y}+V_0)\eta_{,y}-\phi_{,z} =0, \label{eq:KBC}\\
[\mathrm {DBC}]: \quad &  \phi_{,t}+\frac{1}{2}[(\phi_{,x})^2+(\phi_{,y})^2+(\phi_{,z})^2]+U_0 \phi_{,x}+V_0 \phi_{,y}+g\eta =0\label{eq:DBC},
\end{align}
\end{subequations}
where $g$ denotes gravitational acceleration. { In order to apply the boundary conditions, the velocity potential at $z=-H+\eta_b$ and $z=\eta$ need to be respectively Taylor expanded about $z=-H$ and $z=0$.} 
Furthermore, throughout the paper we consider a  {wavy} seabed of the form
\begin{equation}
\eta_b={a_b}\cos{(k_b x + l_b y)},
\label{eq:etab_def}
\end{equation}
where $k_b$ and $l_b$ are respectively the wavenumbers in the $x$- and $y$-directions, $a_b$ is the amplitude,  and $K_b \equiv \sqrt{k_b^2+l_b^2}$.

Hereafter we assume $U_0 \ll V_0$ (and further assume   {$U_0 \lesssim \mathrm{O}(\epsilon^2) $}), since away from the inlets or river mouths, cross-shelf flows are typically much weaker than alongshore flows  \citep{gelfenbaum2005coastal}. We also consider two (small) spatial scales: wave steepness, $\epsilon= K a {\ll 1}$ ($K$ is surface wave's wavenumber and $a$ is its amplitude) and  {wavy} seabed's steepness, $\epsilon_b=K_b a_b {\ll 1}$, and expand the velocity potential ($\phi$) and surface elevation ($\eta$) as perturbation series in terms of $\epsilon$ and $\epsilon_b$. 
The velocity potential and surface elevation can then be expressed in terms of $\epsilon$ and $\epsilon_b$ as
{\begin{subequations}
\begin{align}
\phi&=\big[  \phi_u^{(1)}+  \mathrm{O}(\epsilon^2) \big]+\big[ \phi_s^{(1)}+ \mathrm{O}(\epsilon_b^2)\big], \label{eq: phi_full}\\
\eta&=\big[ \eta_u^{(1)}+\mathrm{O}(\epsilon^2)\big]+\big[ \eta_s^{(1)}+ \mathrm{O}(\epsilon_b^2)\big].
\label{eq: eta_full}
\end{align}
\end{subequations}}
\noindent Both $\phi$ and $\eta$ are a combination of an unsteady solution, denoted by subscript `$u$', and a steady solution, denoted by subscript `$s$' \citep{kirby1988current,fan2021upstream}. {The quantities $\phi_u^{(1)}$ and $\eta_u^{(1)}$ are $\mathrm{O}(\epsilon)$, while $\phi_s^{(1)}$ and $\eta_s^{(1)}$ are $\mathrm{O}(\epsilon_b)$. }  {The steady solution, arising from the interactions between the uniform current and  sinusoidal bottom topography,  manifests itself as stationary waves \citep{thomson1886lx,lamb1932}. Such spatially varying stationary features can also be viewed as $\mathrm{O}(\epsilon_b)$ corrections to the leading order uniform flow due to the wavy seabed topography. }
In \eqref{eq: phi_full}--\eqref{eq: eta_full}, the relationship between $\epsilon$ and $\epsilon_b$ are not yet established, hence they are separated by squared brackets. In the following sections, we will investigate different situations depending on the relationship between $\epsilon$ and $\epsilon_b$.  

\begin{figure}
  \centering
  \begin{subfigure}{0.32\textwidth}
  \centering
  \includegraphics[width=\linewidth]{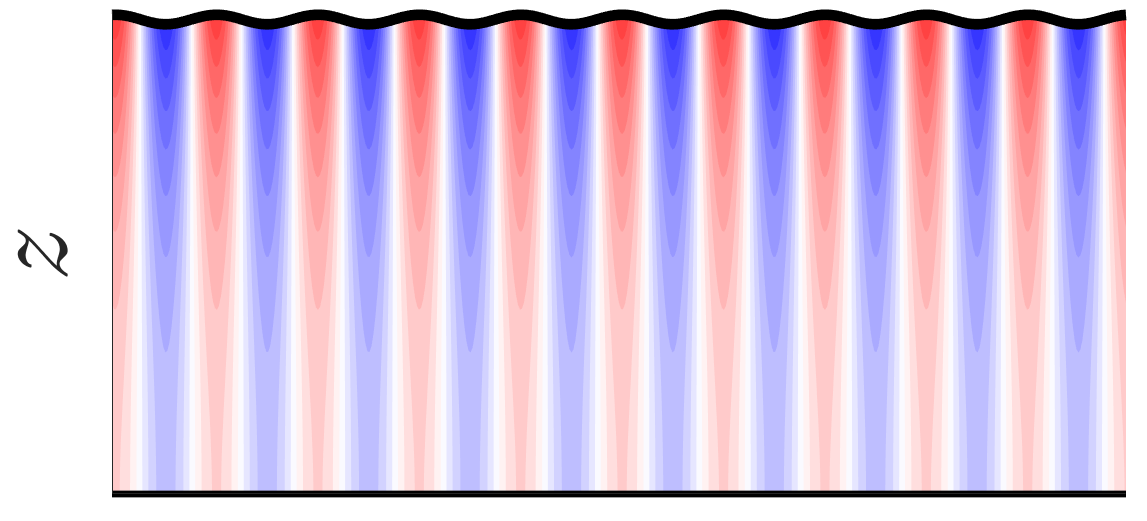}
  \vspace{-3mm}
  \caption{} 
  \end{subfigure}
  \begin{subfigure}{0.295\textwidth}
  \centering
  \includegraphics[width=\linewidth]{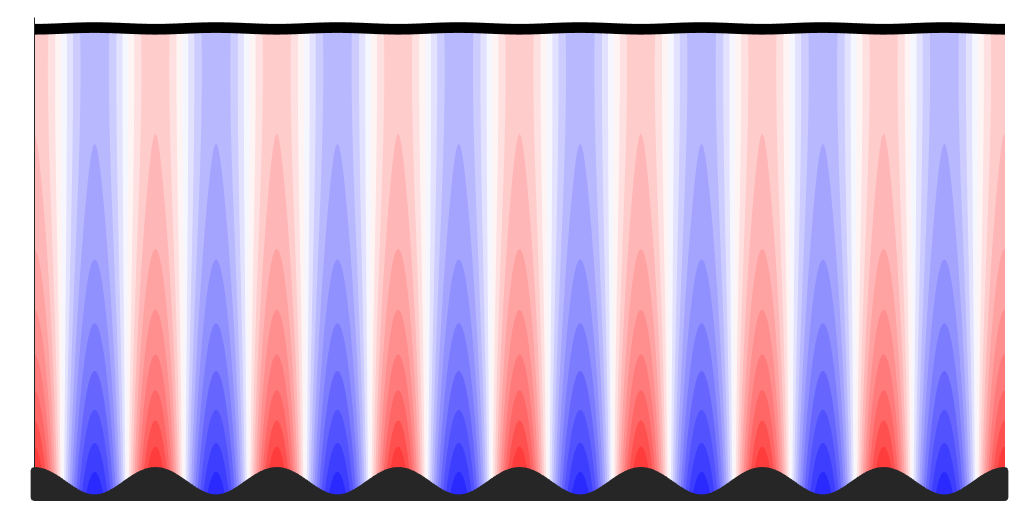}
  \vspace{-3mm}  
  \caption{} 
  \end{subfigure}
  \begin{subfigure}{0.355\textwidth}
  \centering
  \includegraphics[width=\linewidth]{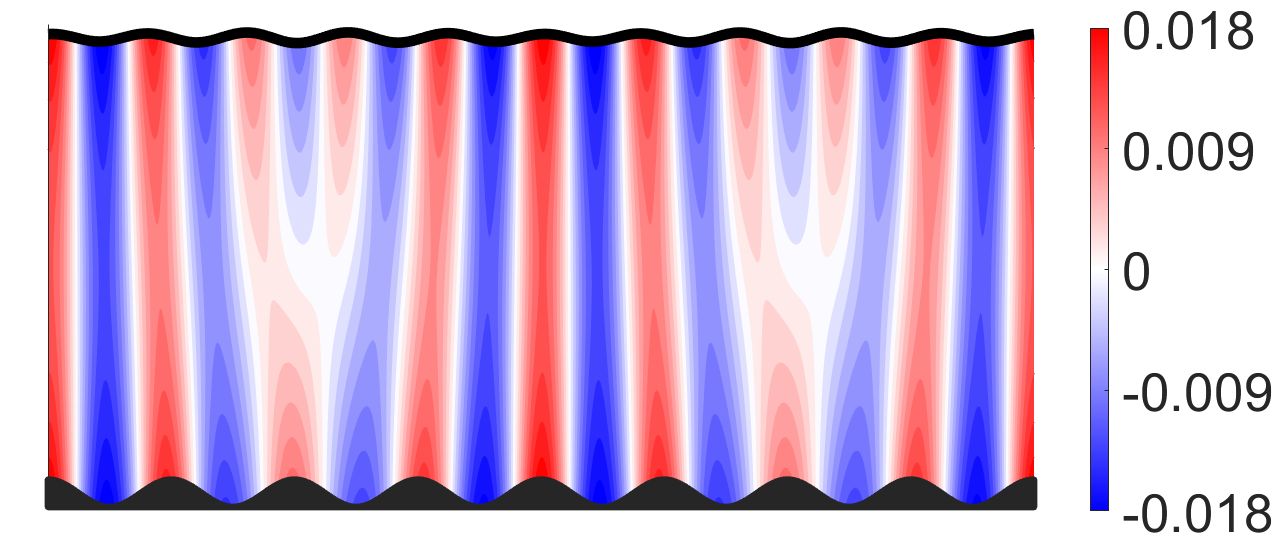}
  \vspace{-3mm}
  \caption{} 
  \end{subfigure}
  \vspace{2mm}
  \begin{subfigure}{0.31\textwidth}
  \centering
  \includegraphics[width=\linewidth]{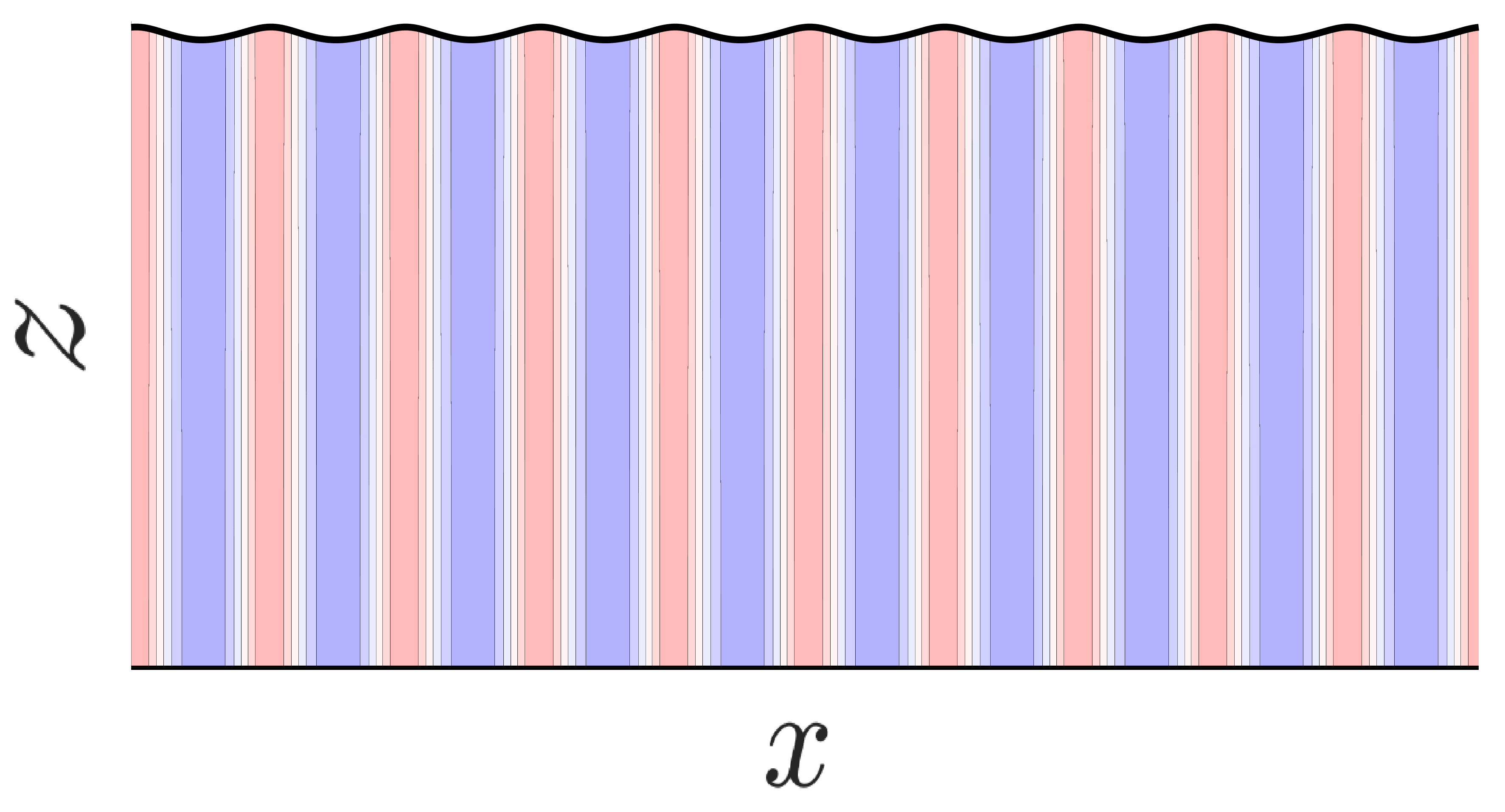}
  \vspace{-3mm}  
  \caption{} 
  \end{subfigure}
  \begin{subfigure}{0.3\textwidth}
  \centering
  \includegraphics[width=\linewidth]{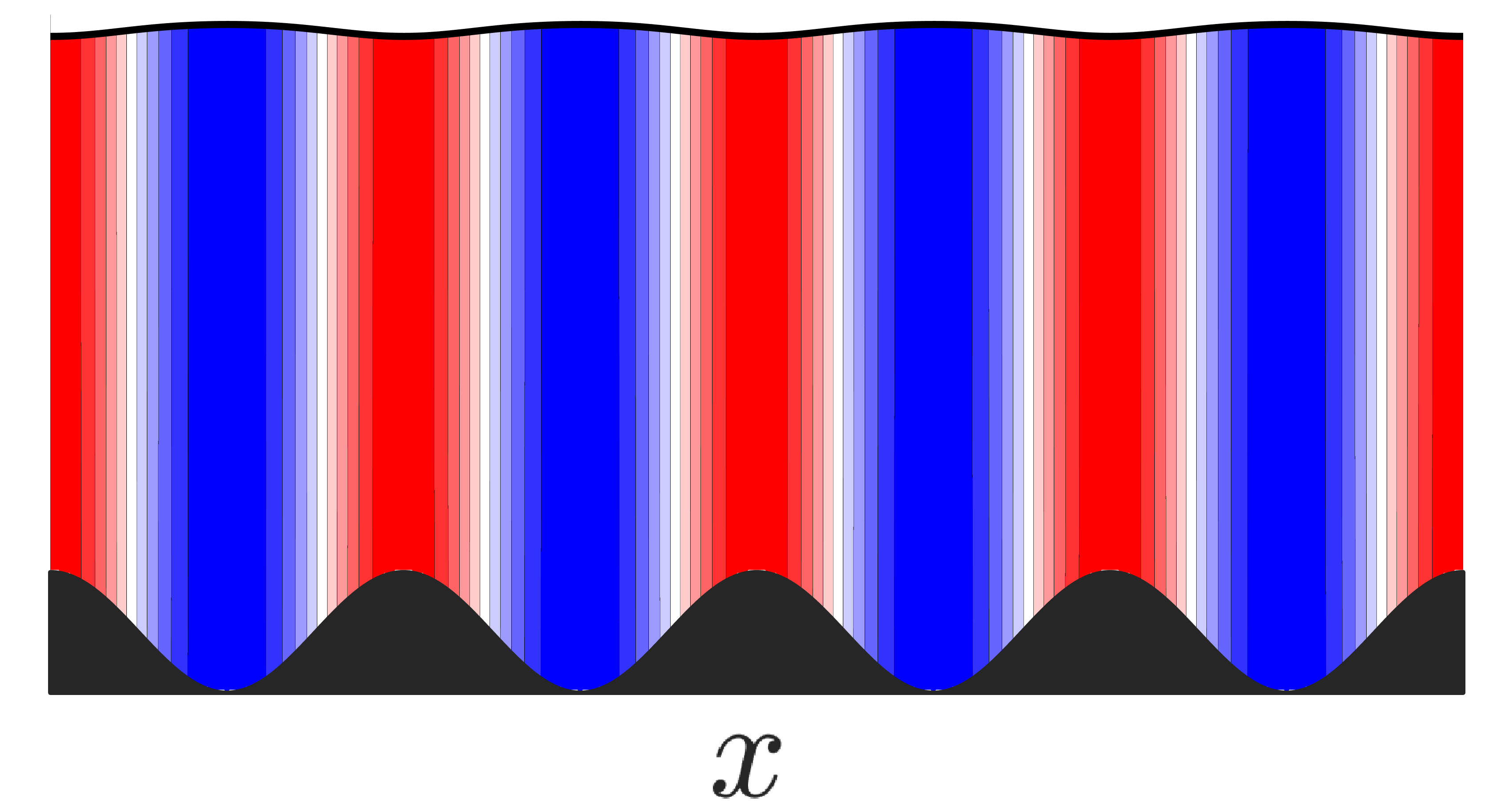}
  \vspace{-3mm}
  \caption{} 
  \end{subfigure}
  \hspace{0.3mm}
  \begin{subfigure}{0.345\textwidth}
  \centering
  \includegraphics[width=\linewidth]{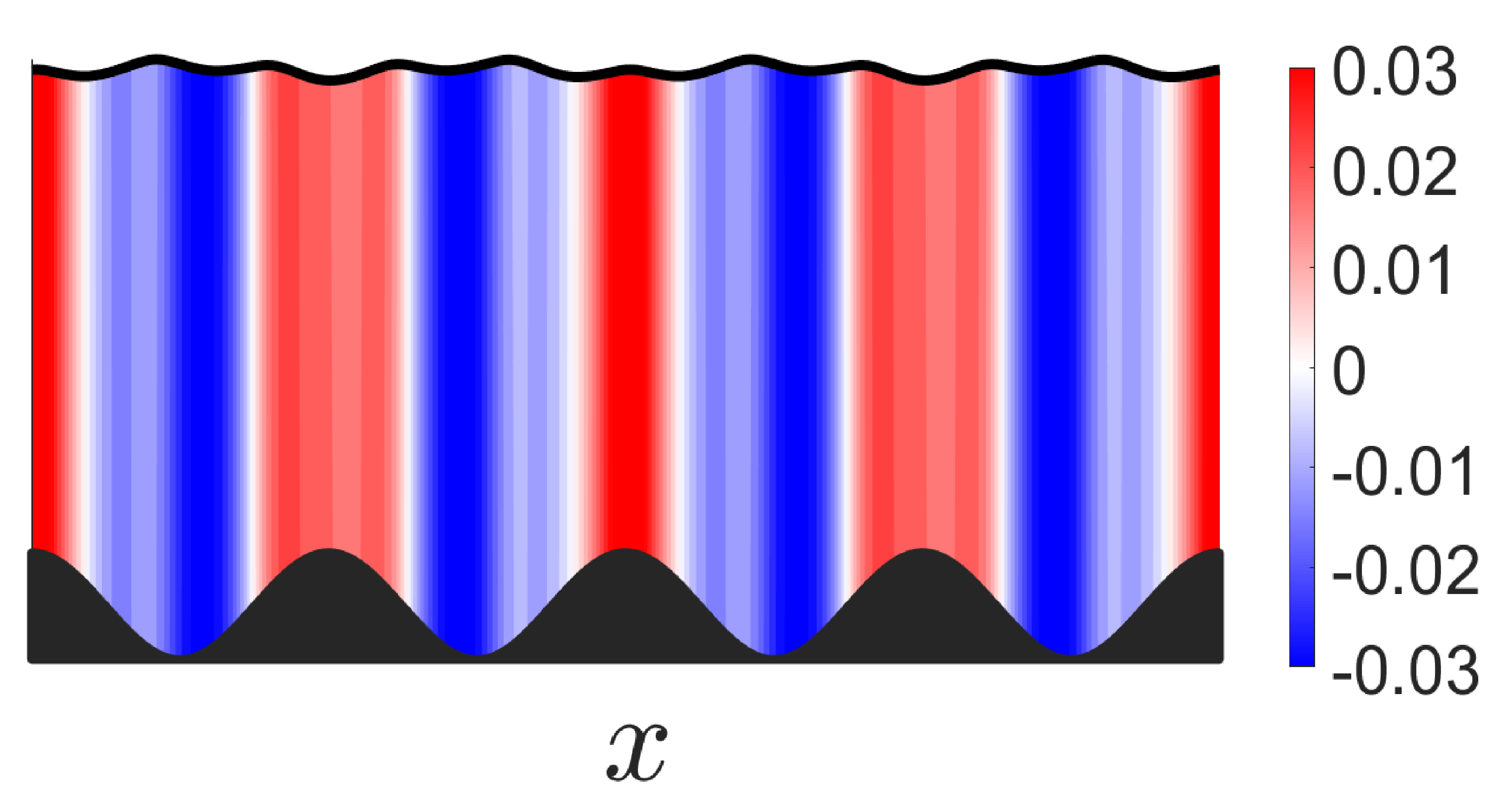}
  \vspace{-3mm}  
  \caption{} 
  \end{subfigure}
  \caption{\footnotesize Contour plots of instantaneous  cross-shelf velocity $u$ in the $x$--$z$ plane for intermediate (a, b, and c) and shallow (d, e, and f) water depths. (a,d): case-I ($\mathrm{O}(\epsilon_b) \ll \mathrm{O}(\epsilon)$), (b,e): case-II ($\mathrm{O}(\epsilon_b) \gg \mathrm{O}(\epsilon)$), and (c,f): case-III ($\mathrm{O}(\epsilon_b)\sim \mathrm{O}(\epsilon)$). Parameters used:  (a) $KH (kH, lH)= 2 (2,0)$, $a/H=0.01$, (b) $K_bH (k_bH, l_bH)= 2 (1.6,1.2)$, $a_b/H=0.03$, (c) combined parameters of (a) and (b), (d) $KH (kH, lH)= 0.2 (0.2,0)$, $a/H=0.01$, (e) $K_bH (k_bH, l_bH)= 0.1 (0.08,0.06)$, $a_b/H=0.1$, and (f) combined parameters of (d) and (e). $Fr\equiv |V_0|/\sqrt{gH}=0.5$ in all cases.}    
  \label{fig:contour_plots}
\end{figure}

The  velocity field can be straightforwardly obtained from the  velocity potential: $\vect{u}\equiv (u,v,w)=(\phi_{,x},\phi_{,y},\phi_{,z})$. Out of the three components, the $u$ velocity  plays the most crucial {role} in the cross-shelf transport of particles.  Figure \ref{fig:contour_plots} shows contour plots of $u$   at an arbitrary time in the $x$--$z$ plane for  different situations depending on the relation between $\epsilon$ and $\epsilon_b$ (these situations have been schematically depicted in figure \ref{fig:Schmatic_diagram}\,{\it b\/}--{\it d\/}). 

The primary focus of this paper is to obtain the trajectory $(x(t),y(t),z(t))$ of a tracer particle, which can be obtained by solving the pathline equations:
\begin{equation}
    \frac{d x}{d t}={U_0 +} u(x,y,z,t), \quad \frac{d y}{d t}=V_0+v(x,y,z,t), \quad \frac{d z}{d t}=w(x,y,z,t).
    \label{eq:pathline}
\end{equation}
Tracer trajectories for the different cases, whose overviews are given in figures \ref{fig:Schmatic_diagram} and \ref{fig:contour_plots}, will be  discussed in the following sections. Unless otherwise mentioned, all tracer trajectories will be studied for particles \emph{at} the free surface.

\vspace{2mm}

{\section{Case-I: Wave steepness dominates over {wavy} seabed's steepness \,\,\,\, [$  \mathrm{O}(\epsilon_b) \ll \mathrm{O}(\epsilon) \,\, {\ll 1} $]}
\label{sec:Case1}}

Here we consider the situation depicted in figure \ref{fig:Schmatic_diagram}\,({\it c\/}) where the wave steepness ($\epsilon$) is much greater than the  {wavy} seabed's steepness ($\epsilon_b)$; i.e. $ \mathrm{O}(\epsilon_b) \ll \mathrm{O}(\epsilon) \,\, {\ll 1}$. In this situation, the bottom surface is perceived to be (nearly) flat. To study the unsteady wave motion over a finite (and constant) depth fluid $H$ and constant alongshore current $V_0$, we substitute the perturbation series of $\phi$ and $\eta$ from \eqref{eq: phi_full}--\eqref{eq: eta_full} into the GLE and BCs, given in \eqref{eq:ImC}, \eqref{eq:KBC}--\eqref{eq:DBC}. At $\mathrm{O}(\epsilon)$ we find:
\begin{subequations}
\begin{align}
[\mathrm {GLE}]:\quad & \phi_{u,xx}^{(1)}+\phi_{u,yy}^{(1)}+\phi_{u,zz}^{(1)}=0 \qquad  & -H  < z < 0, \label{GE_case1}\\
[\mathrm {ImC}]:\quad & \phi_{u,z}^{(1)}=0 \qquad &\mathrm{at} \, z=-H, \label{ImC_case1}\\
[\mathrm {KBC}]:\quad & \eta_{u,t}^{(1)}+V_0\eta_{u,y}^{(1)}-\phi_{u,z}^{(1)}=0  \qquad &\mathrm{at} \, z=0, \label{KBC_case1}\\
[\mathrm {DBC}]:\quad & \phi_{u,t}^{(1)}+V_0 \phi_{u,y}^{(1)}+g\eta_u^{(1)}=0 \qquad &\mathrm{at} \, z=0. \label{DBC_case1}
\end{align}
\end{subequations}
  We assume a linear, progressive {surface} wave of the form
\begin{equation}
    \eta_u^{(1)} = a \cos(k x + l y -\omega t), \label{eq:eta_case1}
\end{equation}
and solve \eqref{GE_case1}--\eqref{DBC_case1}, yielding
\begin{equation}
    \phi_u^{(1)} = \dfrac{a  \overline{\omega}}{K}\dfrac{\cosh K(z+H)}{\sinh(K H)}  \sin(k x + l y -\omega t). \label{eq:phi_case1} 
\end{equation}
Here $a$ is the amplitude and  $\omega$ is the frequency of the surface gravity wave with wavenumber $K$, $\omega=\overline{\omega}+V_0 l$ and $\overline{\omega}(=\sqrt{g K \tanh{(KH)}})$ is its intrinsic frequency, and $k$ and $l$ are respectively the components of $K$ in the $x$- and $y$-directions. It is to be noted that for simplicity, schematic figure \ref{fig:Schmatic_diagram}\,({\it a\/}--{\it c\/}) show surface wave-vector only along $x$.  The  solution \eqref{eq:eta_case1}--\eqref{eq:phi_case1} is generally known as the \emph{homogeneous (unsteady) solution} of  progressive surface gravity waves in the presence of a constant background current in a constant water depth. 
Related (but not the same) set-ups have been thoroughly studied in \cite{peregrine1976interaction,dommermuth1987high,kirby1988current,raj_guha_2019,gupta2021modified}. 

From \eqref{eq:phi_case1}, the  velocity field at $\mathrm{O}(\epsilon)$ can be straightforwardly obtained: $\boldsymbol{u^{(1)}}=\nabla \phi_{u}^{(1)}$. Contours of the $x$--component of $\boldsymbol{u^{(1)}}$ (i.e. $u^{(1)}$) at an arbitrary time have been respectively plotted in figures \ref{fig:contour_plots}\,({\it a\/}) and \ref{fig:contour_plots}\,({\it d\/}) for intermediate ($KH \approx 1$) and shallow ($KH \ll 1$) depths. 

\vspace{7mm}
\subsection{Pathline equations}

\begin{figure}
  \centering
  \begin{subfigure}{0.45\textwidth}
  \centering
  \includegraphics[width=\linewidth]{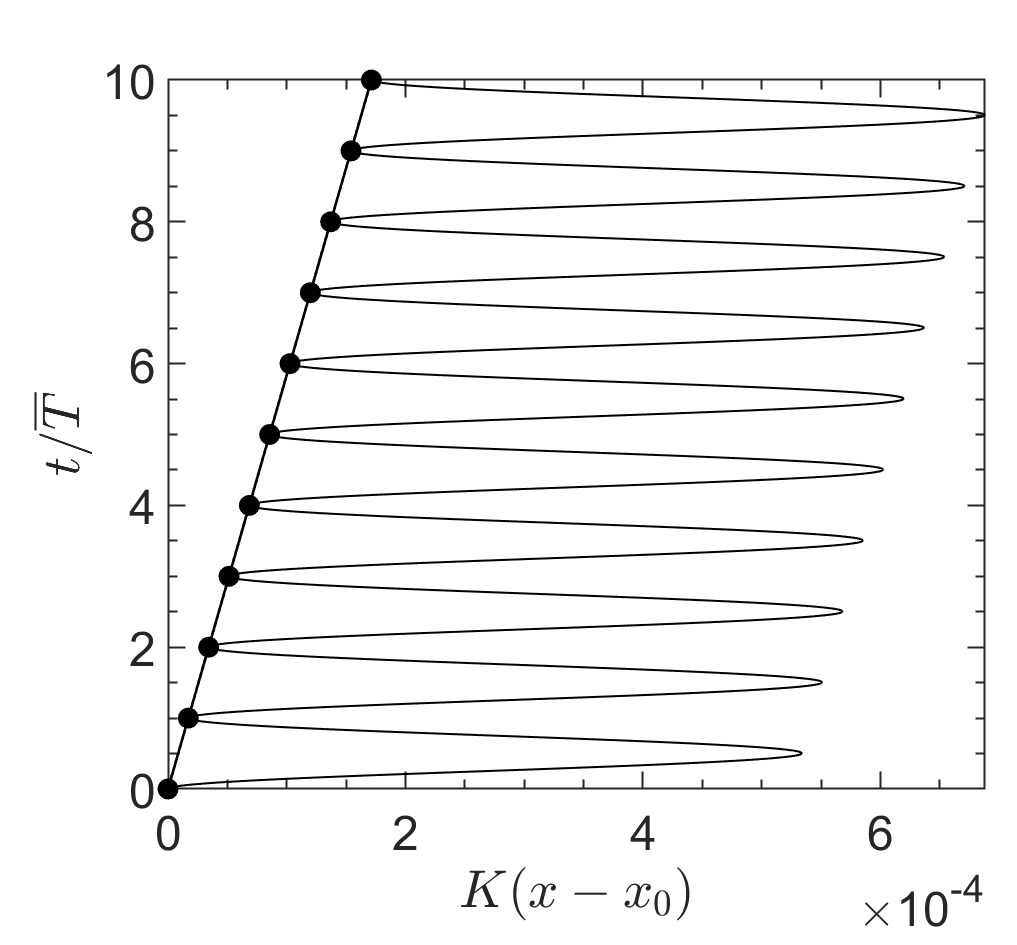}
  \caption{} \label{fig:CaseI}
  \end{subfigure}
   \hspace{5mm}
  \begin{subfigure}{0.45\textwidth}
  \centering
  \includegraphics[width=\linewidth]{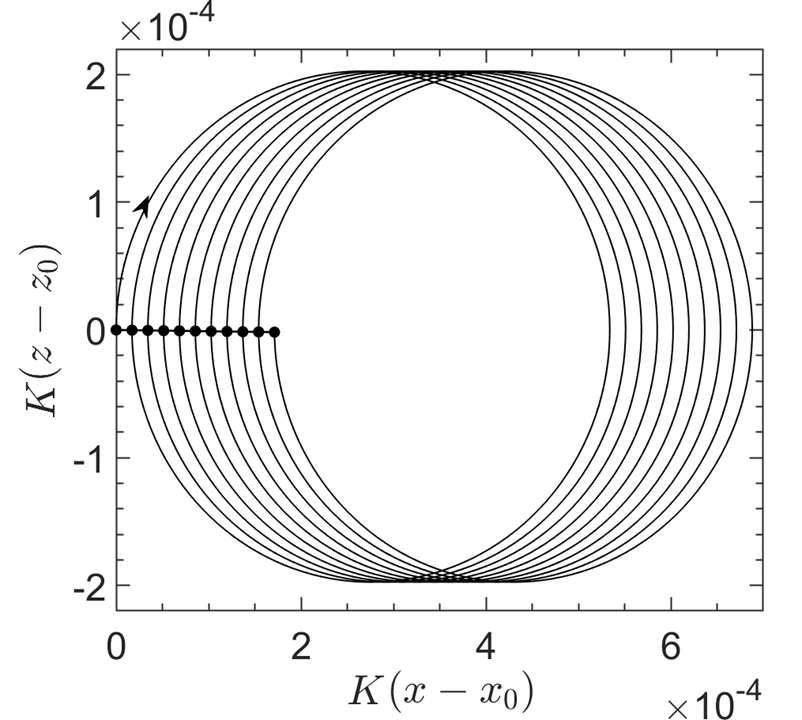}
  \caption{} \label{fig:CaseI_3D}
  \end{subfigure}
  \caption{\footnotesize Particle trajectory for case-I 
  in the non-dimensional  (a) $x$--$t$ plane and (b)  $x$--$z$ plane. These are plotted in a reference frame moving with alongshore current, $V_0$.  The solid, wavy black line denotes the particle trajectory, while filled black circles are plotted after each $\overline{T}$. The black line connecting the circles is the Lagrangian mean trajectory of the particle. Parameters used: $K H (k H, l H)=1 (1, 0)$, $a/H=0.01$, $a_b/H=0$, and $Fr=0.1 \, (V_0>0)$.}   
  \label{fig:CaseI_particle_motion}
\end{figure}

Substitution of $\boldsymbol{u^{(1)}}$ into \eqref{eq:pathline} leads to the pathline equations for case-I:
\begin{subequations}
\begin{align}
 \frac{d x}{d t}  &= \dfrac{a \overline{\omega} k}{K}\dfrac{\cosh K(z+H)}{\sinh (KH)}  \cos \theta , \label{eq:pathline_x_case1}\\
 \frac{d y}{d t} &=V_0+\dfrac{a \overline{\omega} l}{K}\dfrac{\cosh K(z+H)}{\sinh (KH)}  \cos\theta , \label{eq:pathline_y_case1}\\
 \frac{d z}{d t} & = a \overline{\omega}\dfrac{\sinh K(z+H)}{\sinh (KH)}  \sin \theta,
 \label{eq:pathline_z_case1}
\end{align}
\end{subequations}
where $\theta=kx+ly-\omega t$. Even for a linear water wave, a tracer particle moves in an open trajectory, 
{which can be shown by applying phase plane analysis to the nonlinear dynamical system}  \eqref{eq:pathline_x_case1}--\eqref{eq:pathline_z_case1} \citep{henry2007particle, constantin2008particle,constantin2008particle2}. {Explicit solution of this dynamical system is not possible, hence the system needs to be numerically solved in order to evaluate the drift.}
Note that the alongshore current $V_0$  Doppler shifts the frequency, but otherwise does not impact the cross-shelf transport. The forward drift in figure \ref{fig:CaseI_particle_motion}\,({\it a\/}), which is the well-known \emph{Stokes drift}, would lead to  onshore particle transport.  {Hence, linear waves do experience Stokes' drift, which
 is clearly evidenced when the RHS of \eqref{eq:pathline_x_case1}--\eqref{eq:pathline_z_case1} is expanded via small-excursion approximation to include second-order nonlinear terms, given below in \S \ref{subsec:SD}.}

\subsection{Stokes drift: small-excursion approximation in the presence of a uniform current}
\label{subsec:SD}

The classical technique for calculating  Stokes drift employs Taylor expansion by \emph{a-priori} assuming that the particle excursion over one wave-period is small \citep{stokes1847,kundu20fluid,van2017stokes}. Equations   \eqref{eq:pathline_x_case1}--\eqref{eq:pathline_z_case1} include a constant current $V_0$, hence for conducting small-excursion analysis about an initial particle location $(x_0, y_0, z_0)$, we first need to apply the following Galilean transformation: $(X,Y,Z)=(x,y-V_0t,z)$. { This is because in a rest frame, there is always a displacement $V_0t$ along $y$ due to constant advection, which can lead to a violation of the small-excursion approximation.} {A} closed {elliptical} trajectory is obtained at $\mathrm{O}(\epsilon)$, but inclusion of $\mathrm{O}(\epsilon^2)$ terms reveal an open trajectory, {shown in figure \ref{fig:CaseI_particle_motion}\,({\it b\/})}, and yields the Stokes drift velocity:
\begin{subequations}
\begin{align}
\langle u_{SD} \rangle &=\frac{a^2 \overline{\omega} k}{2  \sinh^2{(KH)}}\cosh[2K(z_0+H)], \label{eq:u_sd}\\
\langle v_{SD} \rangle &=\frac{a^2 \overline{\omega} l}{2  \sinh^2{(KH)}}\cosh[2K(z_0+H)], \label{eq:v_sd}\\
\langle w_{SD} \rangle &= 0, \label{eq:w_sd}
\end{align}
\end{subequations}
where $\langle \ldots \rangle$ denotes averaging over one wave period (in the moving frame), $\overline{T}=2\pi/\overline{\omega}$. Equations analogous to \eqref{eq:u_sd}--\eqref{eq:w_sd} can  be found in  \cite{ursell1953long,gupta2021modified}.
The small-excursion approximation, which serves the basis for \eqref{eq:u_sd}--\eqref{eq:w_sd}, yields a highly accurate  solution --  {for deep and shallow water waves, the errors are respectively  $\mathrm{O}(\epsilon^6)$  \citep{longuet1987lagrangian,van2017stokes} and $\mathrm{O}(\epsilon^4)$ \citep{clamond2007lagrangian}}. { The resulting Stokes drift displacement is 
$$\boldsymbol{\Delta x_{SD}}= \langle \boldsymbol{u_{SD}}\rangle \, \overline{T},$$
and is the linear distance between two filled black circles shown in figure \ref{fig:CaseI_particle_motion}, where  $\overline{T}=T$ since $l=0$.}
 
Among other things, we will scrutinize in \S \ref{sec:Case2}  whether \emph{a-priori} assumption of small-excursion applied to the pathline equations yields highly accurate results when there is a  {wavy} seabed and a background alongshore current.

\subsection{Lagrangian drift and Eulerian return flow}
\label{sec:3.3}
In a typical nearshore environment,  predominantly onshore propagating surface waves lead to an onshoreward  Stokes drift velocity. However, due to the presence of shoreline, there is a compensating wave-driven offshore flow, generally referred to as the Eulerian return flow or undertow, $U_{E(SD)}(z)$ \citep{lentz2008observations, brown2015field}. While the depth integrated Stokes drift velocity balances the   depth integrated Eulerian return flow,  they do not necessarily cancel at any particular depth \citep{longuet1953mass,kumar2017effect}.   The resulting cross-shelf Lagrangian drift velocity, $U_{L(SD)}$ at any depth can be  written as: 
\begin{equation}
  U_{L(SD)}(z) =U_{E(SD)}(z)+\langle u_{SD}(z) \rangle.
  \label{eq:LagDri_caseI}
\end{equation}
$U_{E(SD)}$ profile varies from the surf zone to the inner shelf,  and is usually reconstructed from observational studies \citep{lentz2008observations}. In the conceptual model outlined in \S \ref{sec:formulation},  { the Eulerian return flow can be represented (albeit in  a simplified way) by the weak, uniform cross-shelf current $U_0$ by assuming that the net horizontal mass transport is zero, i.e.
\begin{equation}
    U_{E(SD)}=U_0=-\frac{1}{H}\int_{-H}^{0} \langle u_{SD}(z) \rangle dz=-\frac{a^2 \overline{\omega} k}{2KH} \coth{KH}.
\end{equation}
The quantity $\int_{-H}^{0} \langle u_{SD}(z) \rangle dz$  is  known as the \emph{Stokes transport}, which is a measure of the depth integrated mass transport by surface waves.} {$U_0$ can be written in terms of Stokes drift velocity at the free surface:
\begin{equation}
    \frac{U_0}{\langle u_{SD}(z=0) \rangle}=-\frac{\tanh (2KH)}{2KH}. \label{eq:U0_uSD}
\end{equation}
Hence in the shallow water-limit,
$U_0=-\langle u_{SD} \rangle$, resulting in $U_{L(SD)}=0$, while in the deep water-limit, $U_0\ll -\langle u_{SD} (z=0)\rangle$.}

Although particle trajectories in figure \ref{fig:CaseI_particle_motion} do not account for the Eulerian return flow, it could be included  by simply adding $U_0$ in the  pathline equation along the $x$-direction. In the subsequent sections, we will investigate whether alongshore current and  {wavy} seabed interactions could act as an additional cross-shelf transport mechanism.

\vspace{-4mm}

{\section{Case II:  {Wavy} seabed's steepness dominates over wave steepness  \,\, [$\mathrm{O}(\epsilon) \ll \mathrm{O}(\epsilon_b) {\ll 1}$]}
\label{sec:Case2}}

Here we consider the situation shown in figure \ref{fig:Schmatic_diagram}\,({\it d\/}) -- there is a uniform alongshore current ($V_0$) over a  {wavy} bottom topography ($\eta_b$), but surface waves are either absent or have negligible effects (mathematically, $\mathrm{O}(\epsilon) \ll \mathrm{O}(\epsilon_b) {\ll 1}$). To study this, we substitute the perturbation series of $\phi$ and $\eta$ from \eqref{eq: phi_full}--\eqref{eq: eta_full} into the GLE and BCs, given in \eqref{eq:ImC}, \eqref{eq:KBC}--\eqref{eq:DBC}. At $\mathrm{O}(\epsilon_b)$, we obtain the following set of steady, non-homogeneous system of equations: 
\begin{subequations}
\begin{align}
[\mathrm {GLE}]:\quad & \phi_{s,xx}^{(1)} +\phi_{s,yy}^{(1)} +\phi_{s,zz}^{(1)} =0 \qquad & -H < z < 0 \label{eq:GE_case2},\\
[\mathrm {ImC}]:\quad & \phi_{s,z}^{(1)} =V_0 \eta_{b,y}  \qquad &\mathrm{at} \, z=-H, \label{eq:ImC_case2}\\
[\mathrm {KBC}]:\quad &  V_0\eta_{s,y}^{(1)} - \phi_{s,z}^{(1)} = 0  \qquad &\mathrm{at} \, z=0, \label{eq:KBC_case2} \\
[\mathrm {DBC}]:\quad &  V_0 \phi_{s,y}^{(1)} +g\eta_s^{(1)} =0 \qquad &\mathrm{at} \, z=0. \label{eq:DBC_case2}
\end{align}
\end{subequations}
The steady surface elevation and  velocity potential, which results from the interaction between the  {wavy} bottom boundary and the uniform alongshore current, and obtained by solving \eqref{eq:GE_case2}--\eqref{eq:DBC_case2} along with \eqref{eq:etab_def}, are respectively given by:
\begin{subequations}
\begin{align}
\eta_s^{(1)}(x,y)=& \, {a_{s}} \cos{(k_b x+l_b y)}, \qquad \mathrm{and}\label{eq:eta_case2}\\
\phi_s^{(1)}(x,y,z)=&  \bigg[  {A_{s}} \frac{\cosh K_b(z+H)}{\cosh(K_b H)}+ {B_{s}} \frac{\sinh(K_b z)}{\cosh(K_b H)} \bigg] \sin{ (k_b x+l_b y)}. \label{eq:phi_case2}
\end{align}
\end{subequations}
Here,\begin{align*}
    a_{s}&={\dfrac{V_0^2 l_b^2 a_b }{[V_0^2 l_b^2- g K_b \tanh(K_b H)] \cosh(K_b H)}}, \\
    A_{s}&=-\dfrac{  V_0 l_b g a_b}{[V_0^2 l_b^2- g K_b \tanh(K_b H)] \cosh(K_b H)}, \textnormal{and}\\
    B_{s}&=-\dfrac{ V_0 l_b a_b }{K_b}. 
\end{align*} 
The expressions \eqref{eq:eta_case2}--\eqref{eq:phi_case2} denote the \emph{particular (steady) solution}, { and have been previously investigated by many authors in various contexts \citep{lamb1932,kennedy1963mechanics,sammarco1994nonlinear,kirby1988current,fan2021upstream}.}  Equations \eqref{eq:eta_case2}--\eqref{eq:phi_case2}  result from the presence of the non-homogeneous term $V_0 \eta_{b,y}$ in the R.H.S. of \eqref{eq:ImC_case2}. If either  $\eta_{b,y}=0$ (i.e. no topographic variation in the alongshore direction) or $V_0=0$,  there would be no steady surface impressions, and no steady  velocity potential. We emphasize here that $\eta_{b,y}=0$ in \eqref{eq:etab_def} would imply $a_bl_b=0$. Therefore, sinusoidal  {bottom boundary} ($a_b \neq0$ and $k_b \neq0$) with $l_b=0$ would still lead to a null or trivial particular solution. This is also evident from the dependence of $a_s$, $A_s$, and $B_s$ on $l_b$. 

Figures \ref{fig:contour_plots}\,({\it b\/}) and \ref{fig:contour_plots}\,({\it e\/}) respectively show $u^{(1)}$(=$\phi_{s,x}^{(1)}$, where $\phi_{s}^{(1)}$ is in \eqref{eq:phi_case2}) contours for intermediate and shallow depths. Figure \ref{fig:contour_plots}\,({\it b\/}) reveals an obvious, yet important fact that $|u^{(1)}|$ is maximum at the bottom and decays with elevation, contrary to the behaviour observed in figure \ref{fig:contour_plots}\,({\it a\/})  (the intermediate depth situation for case-I, i.e. the `homogeneous' problem).
Additionally, figures \ref{fig:contour_plots}\,({\it b\/}) and \ref{fig:contour_plots}\,({\it e\/}) reveal a standard result of open-channel hydraulics -- for sub-critical flow, i.e. when the Froude number, $Fr \equiv |V_0|/\sqrt{gH}<1$, the surface impressions are $\pi$ shifted from the bottom undulations.

\vspace{7mm}


\subsection{Pathline equations}
\label{subsec:path_cbiid}

Under the umbrella of the wide range of problems associated with the `water-wave theory', probably the only known pathline equation is \eqref{eq:pathline_x_case1}--\eqref{eq:pathline_z_case1}, or its minor variations (e.g. when background current is absent), yielding the celebrated Stokes drift, i.e. mass transport by surface waves. However, even when surface waves are absent, it is \emph{still} possible to obtain pathline equations. In this case, the  velocity field is obtained \emph{not} from the homogeneous/unsteady solution \emph{but} from the particular/steady solution  $\boldsymbol{u^{(1)}}=\nabla \phi_{s}^{(1)}$. These pathline equations up to $\mathrm{O}(\epsilon_b)$ are given by:

\begin{subequations}
\begin{align}
\frac{d x}{d t} &=  k_b \bigg[  {A_{s}} \frac{\cosh K_b(z+H)}{\cosh(K_b H)}+ B_{s}\frac{\sinh(K_b z)}{\cosh(K_b H)} \bigg] \cos{\theta_b}, \label{eq:pathline_x_case2}\\
\frac{d y}{d t} &=V_0 + l_b \bigg[  {A_{s}} \frac{\cosh K_b(z+H)}{\cosh(K_b H)}+ {B_{s}} \frac{\sinh(K_b z)}{\cosh(K_b H)} \bigg] \cos{\theta_b}, \label{eq:pathline_y_case2}\\
\frac{dz}{d t} &= K_b \bigg[ {A_{s}} \frac{\sinh K_b(z+H)}{\cosh(K_b H)}+ {B_{s}} \frac{\cosh(K_b z)}{\cosh(K_b H)} \bigg] \sin{\theta_b}, \label{eq:pathline_z_case2}
\end{align}
\end{subequations} 
where $\theta_b=k_b x+l_b y$. As already mentioned, $l_b\neq 0$ is necessary for the existence of a non-trivial particular solution. Furthermore, \eqref{eq:pathline_x_case2} reveals
that the pathline equation in the $x$-direction is dependent on $k_b$. Hence, both $k_b\neq 0$ and $l_b \neq 0$ are necessary for the existence of any motion  in the $x$ (i.e. cross-shelf) direction.

\vspace{1mm}
\subsubsection{Small-excursion approximation in the presence of a uniform current}
\label{subsec:4.1.1}

\begin{figure}
    \centering
    \begin{subfigure}{0.48\textwidth}
    \includegraphics[width=\linewidth]{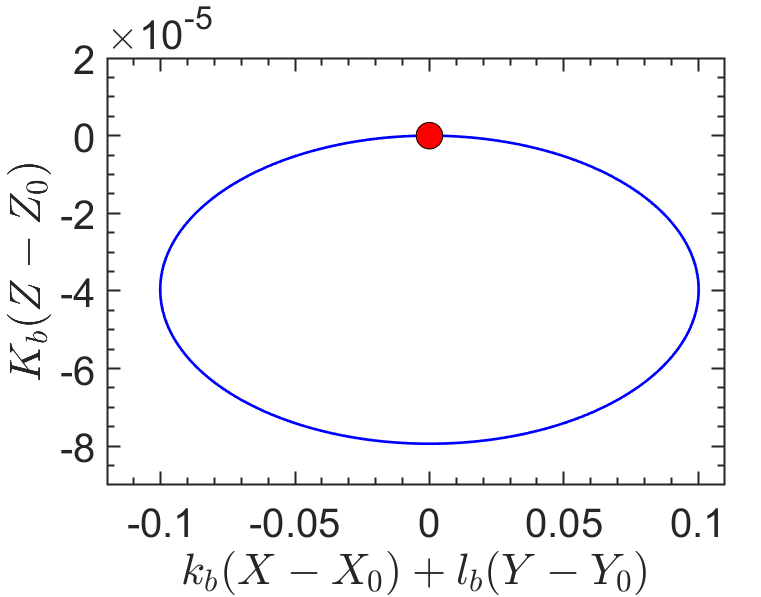}
    \end{subfigure}
    \hspace{3mm}
    \begin{subfigure}{0.48\textwidth}
    \includegraphics[width=\linewidth]{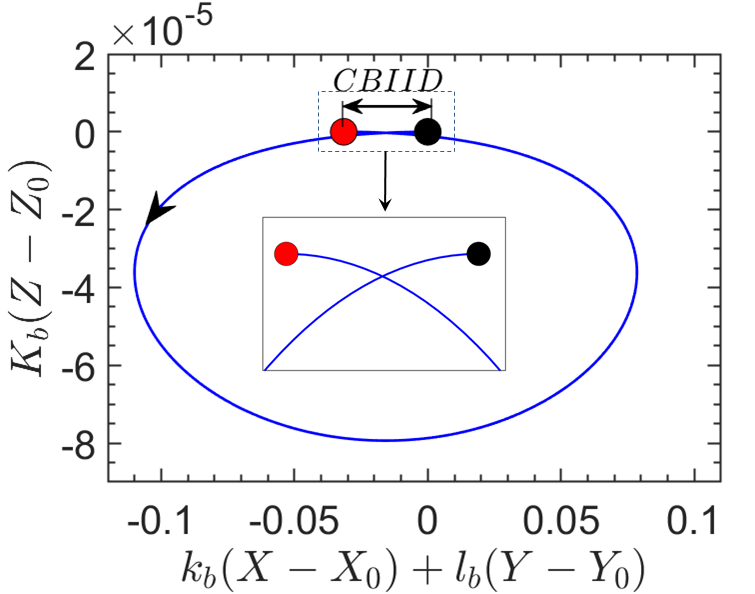}
    \end{subfigure}
    \caption{\footnotesize Particle trajectory in a reference frame moving with the alongshore current, $V_0$. Small-excursion approximation shows (a) closed trajectory at $\mathrm{O}(\epsilon_b)$, and (b) open trajectory up to $\mathrm{O}(\epsilon_b^2)$. The latter reveals CBIID, analogous to Stokes drift by surface waves. Filled black circle denotes initial position while filled red circle denotes  position after one time period. Parameters used: $a_b/H=0.1$,  $K_b H (k_b H, l_b H)=0.1 (0.08,0.06)$, $a/H=0$, $Fr=0.1 \, (V_0>0)$.}
    \label{fig:traj_open_close}
\end{figure}

We follow a procedure similar to what has been outlined in \S \ref{subsec:SD} -- apply Galilean transform  $(X,Y,Z)=(x,y-V_0t,z)$ to \eqref{eq:pathline_x_case2}--\eqref{eq:pathline_z_case2} and Taylor expand about an initial position, assuming that the excursion in one time period is small. {A} closed particle trajectory is obtained at $\mathrm{O}(\epsilon_b)$, analogous to that obtained for surface waves at $\mathrm{O}(\epsilon)$; see figure \ref{fig:traj_open_close}\,({\it a\/}). The locus is the equation of an ellipse:
{
\begin{equation}
    \frac{\{k_b (X-X_0)+ l_b (Y-Y_0)\}^2}{(K_b^2 {\mathbb{P}}/V_0 l_b)^2}+\frac{\{K_b (Z-Z_0)\}^2}{(K_b^2 {\mathbb{Q}}/V_0 l_b)^2}=1,
\end{equation}}
where  
\begin{align*}
 {\mathbb{P}}=\bigg[{A_{s}} \dfrac{\cosh K_b(z_0+H)}{\cosh(K_b H)}+ B_{s}\dfrac{\sinh(K_b z_0)}{\cosh(K_b H)}\bigg],\,\,
{\mathbb{Q}}=\bigg[{A_{s}} \dfrac{\sinh K_b(z_0+H)}{\cosh(K_b H)}+ {B_{s}} \dfrac{\cosh(K_b z_0)}{\cosh(K_b H)}\bigg].
\end{align*}

\noindent{Hence, in the moving frame, particle trajectories are confined to a plane formed by the bottom {topography}  wave-vector, $\vect{K_b}$, and the $z$-axis.}  
{An} open trajectory shown in  figure \ref{fig:traj_open_close}\,({\it b\/}) is obtained at $\mathrm{O}(\epsilon_b^2)$,  analogous to that observed for surface waves at $\mathrm{O}(\epsilon^2)$. We refer to this new kind of drift  as the current-bathymetry interaction induced   drift (CBIID). The \emph{approximate}-CBIID (aCBIID) velocity, obtained using the small-excursion approximation (which is the analog of Stokes drift velocity \eqref{eq:u_sd}--\eqref{eq:w_sd} for surface waves), is given by 
\begin{equation}
\langle \boldsymbol{u_{aCBIID}} \rangle =\langle (\boldsymbol{X}-\boldsymbol{X_0}) \cdot \nabla \boldsymbol{u^{(1)}}|_{\boldsymbol{X=X_0}}\rangle,
\label{eq:formula_uaCBIID}
\end{equation}
where $\boldsymbol{u^{(1)}}$ is evaluated in the moving frame. In component form, this finally yields
\vspace{-2mm}
\begin{subequations}
\begin{align}
\langle u_{aCBIID} \rangle &=-\frac{k_b K_b^2}{2 V_0 l_b} ({\mathbb{P}}^2+{\mathbb{Q}}^2), \label{eq:u_aCBIID}\\
\langle v_{aCBIID} \rangle &=-\frac{l_b K_b^2}{2 V_0 l_b} ({\mathbb{P}}^2+{\mathbb{Q}}^2), \label{eq:v_aCBIID}\\
\langle w_{aCBIID} \rangle &= 0, \label{eq:w_aCBIID}
\end{align}
\end{subequations}
where $\langle \ldots \rangle$ denotes averaging over one time period in the moving frame, $T_{aCBIID}=2\pi/V_0 l_b$.
{Analogous to  Stokes transport (see \S \ref{sec:3.3}), we can introduce `CBIID transport', which is defined as the depth integrated mass transport by steady surface imprints. The CBIID transport in the $x$- and $y$-directions are given respectively as follows:
\begin{subequations}
    \begin{align}
     \int_{-H}^{0} \langle u_{aCBIID}\rangle dz =  - \frac{k_b K_b^2}{2 V_0 l_b}  \Gamma,
     \label{eq:Uretcbiid}\\
     \int_{-H}^{0} \langle v_{aCBIID} \rangle dz = - \frac{l_b K_b^2}{2 V_0 l_b}  \Gamma,
    \end{align}
        \end{subequations}
    where $\Gamma=\dfrac{(A_s^2+B_s^2)\sinh(2K_b H)}{2 K_b \cosh^2(K_bH)}$.}

{The cross-shelf Lagrangian velocity  is given by}
\begin{equation}
  U_{L(CBIID)}(z)=U_{E(CBIID)}(z)+\langle u_{aCBIID}(z) \rangle.
  \label{eq:noreturn}
\end{equation}
{Following \S \ref{sec:3.3}, the Eulerian return flow due to CBIID, $U_{E(CBIID)}$ can be assumed constant as a first approximation, leading to $U_{E(CBIID)}=U_0=k_b K_b^2 \Gamma/(2 V_0 l_b H)$. Moreover, analogous to Stokes drift in the shallow-water limit, $U_0=-\langle u_{aCBIID} \rangle$ in the long-bottom-undulation limit (i.e. $K_bH\ll 1$), resulting in $U_{L(CBIID)}=0$.}

Exact particle trajectories, obtained by solving \eqref{eq:pathline_x_case2}--\eqref{eq:pathline_z_case2},  are plotted in figure \ref{fig:caseII} for different parameter regimes. These figures indeed show that the steady velocity field arising from the particular solution leads to a cross-shelf tracer transport. All configurations in  figure \ref{fig:caseII} have $V_0>0$, leading to a CBIID displacement that is directed towards $x<0$, as evident from \eqref{eq:u_aCBIID}. Figures \ref{fig:caseII}\,({\it a\/}) and \ref{fig:caseII}\,({\it b\/}) show trajectories in the $x$--$t$ plane for the intermediate--depth or {moderate-bottom-undulation} limit ($K_bH\approx 1$), while  figures \ref{fig:caseII}\,({\it c\/}) and \ref{fig:caseII}\,({\it d\/}) show the same for the shallow-water or {long-bottom-undulation} limit ($K_bH\ll 1$). For intermediate-depth and small bottom {topography} height (i.e. $a_b/H \ll 1$), the trajectory obtained from the small-excursion approximation is nearly indistinguishable from that obtained from the exact solution; see figure \ref{fig:caseII}\,({\it a\/}). However, differences arise as the bottom   {topography} height is increased to  $a_b/H=0.1$, see figure \ref{fig:caseII}\,({\it b\/}). For the shallow-water case, differences between  the small-excursion approximation and the exact solution are visible even when bottom   {topography} height is small (figure \ref{fig:caseII}\,{\it c\/}), and the differences get larger with increasing the bottom  {topography} height (figure \ref{fig:caseII}\,{\it d\/}). We note in passing that figure \ref{fig:caseII}\,({\it a--d\/}) shows $t/T_{CBIID}\in [45,50]$; during the initial times (not shown in figure), the small-excursion approximation is nearly indistinguishable from the exact solution. Moreover, for the deep-water or {short-bottom-undulation} case ($K_bH\gg 1$), the small-excursion approximation matches nearly exactly with the exact solution even when $a_b/H=0.1$  (not shown in figure).

\vspace{2mm}
\subsubsection{Near--exact solution: the $z$-bounded approximation}
\label{sec:z_bounded}

\begin{figure}
  \hspace{-5mm}
  \centering
  \begin{subfigure}{0.5\textwidth}
  \centering
  \includegraphics[width=\linewidth]{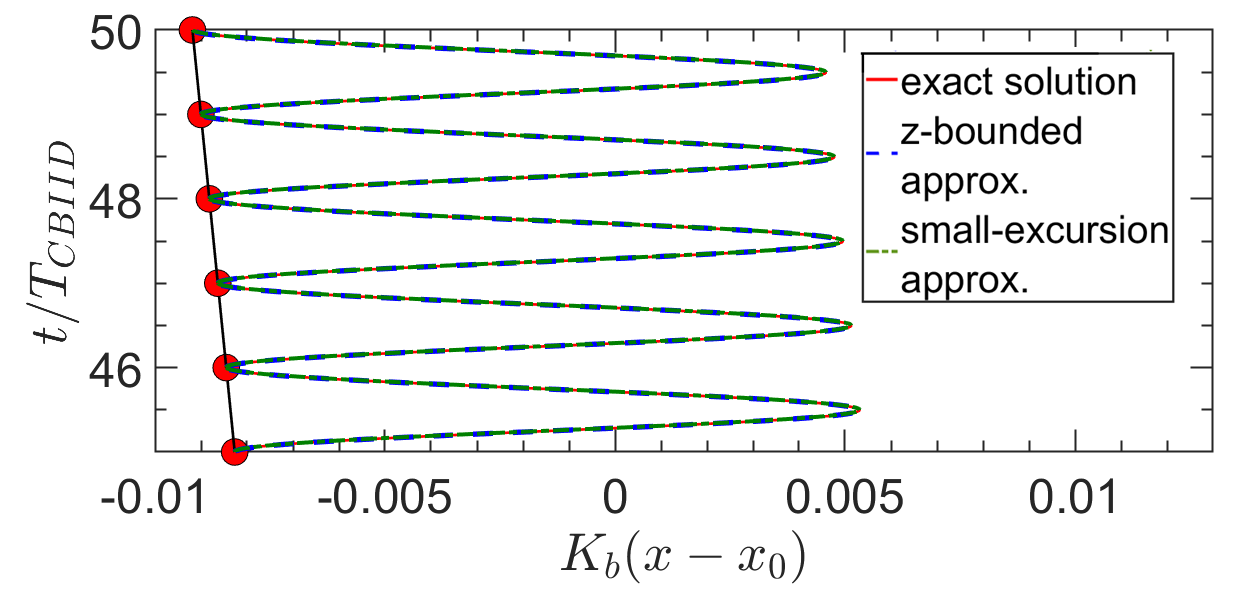}
  \caption{} \label{fig:CaseII_particle_motion_Kb=50_ab=0.01H}
  \end{subfigure}
  \begin{subfigure}{0.5\textwidth}
  \centering
  \includegraphics[width=\linewidth]{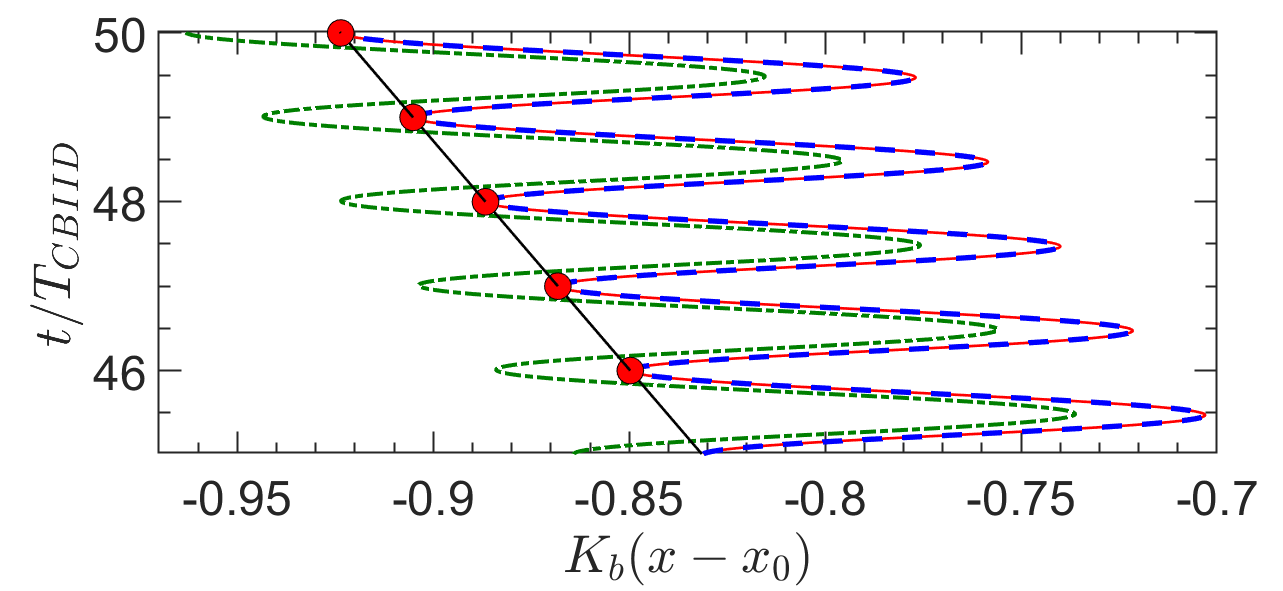}
  \caption{} \label{fig:CaseII_particle_motion_Kb=50_ab=0.1H}
  \end{subfigure}\\
  \hspace{-3mm}
  \begin{subfigure}{0.505\textwidth}
  \centering
  \includegraphics[width=\linewidth]{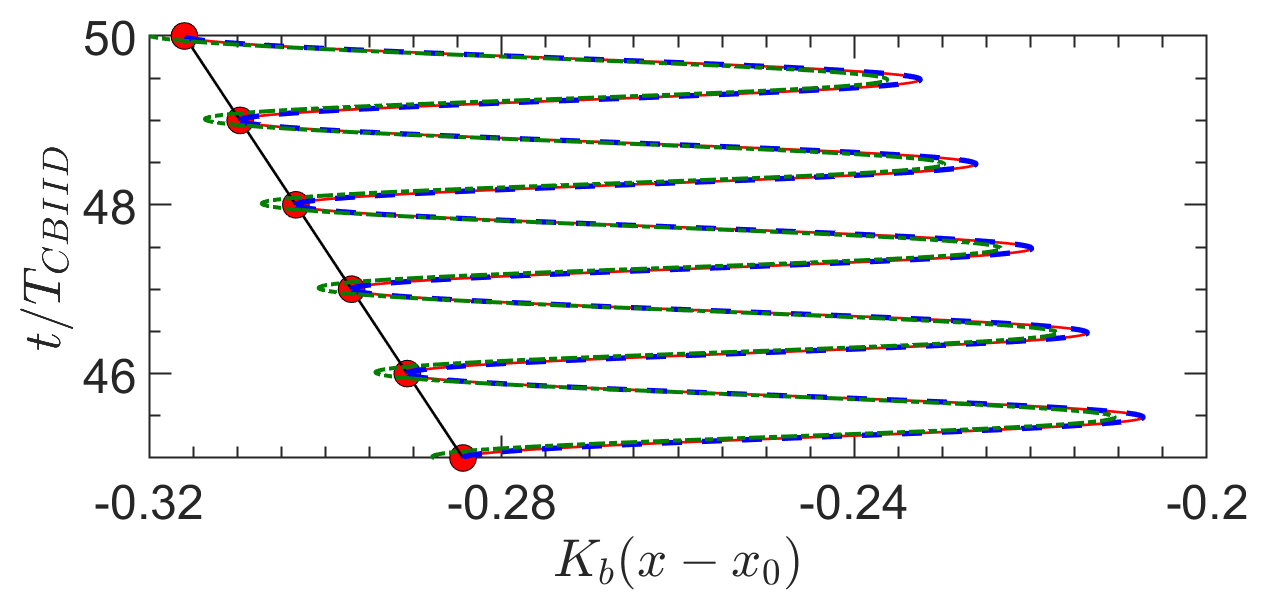}
  \caption{} \label{fig:CaseII_particle_motion_Kb=5_ab=0.05H}
  \end{subfigure}
   \begin{subfigure}{0.5\textwidth}
  \centering
  \includegraphics[width=\linewidth]{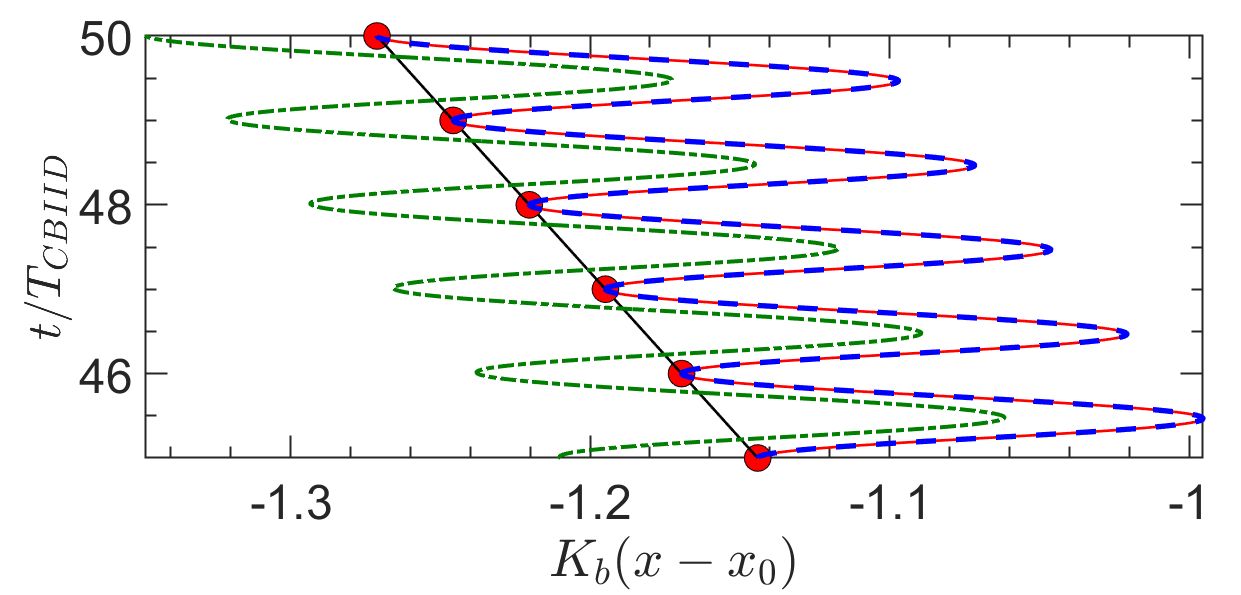}
  \caption{} \label{fig:CaseII_particle_motion_Kb=5_ab=0.1H}
  \end{subfigure}
  \vspace{5mm}
  \begin{subfigure}{0.6\textwidth}
  \centering
  \includegraphics[width=\linewidth]{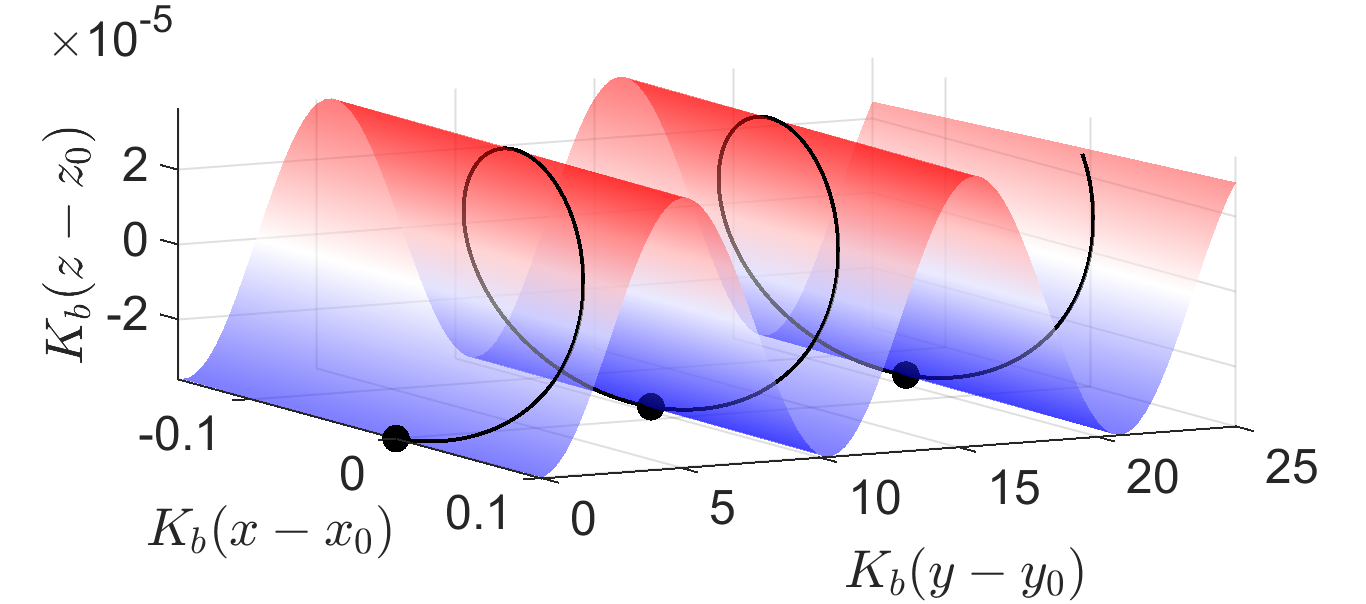}   
  \caption{} \label{fig:surf_traj_caseII}
  \end{subfigure}
  \caption{\footnotesize Particle trajectory for case-II, the non-homogeneous (steady) solution. Particle trajectory in the non-dimensional $x$--$t$ plane for (a, b) intermediate-depth/{moderate-bottom-undulation} with $K_b H (k_b H, l_b H)=1 (0.8,0.6)$, and (c, d) shallow-water/{long-bottom-undulation} with $K_b H (k_b H, l_b H)=0.1 (0.08,0.06)$. Solid red, dashed blue, and dash{-dotted} green curves respectively denote the exact solution, the $z$-bounded approximation, and the small-excursion approximation.  Filled red circles are plotted after each time period, $T_{CBIID}$, and are connected by the Lagrangian mean trajectory (solid black line). (e) Free surface impression, $\eta_s$, is shown by the surface plot for $K_b H (k_b H, l_b H)=0.1 (0.08,0.06)$. Particle trajectory (which is always on the free surface) is shown by the solid black curve, and is plotted for the stationary reference frame. Filled black circles denote positions after each $T_{CBIID}$. Bottom {undulation} heights are as follows: (a) $a_b/H=0.01$, (b) $a_b/H=0.1$, (c) $a_b/H=0.05$, and (d,e) $a_b/H=0.1$. For all cases, $a/H=0$ (no surface wave) and  $Fr=0.1 \, (V_0>0)$.}  
  \label{fig:caseII}
\end{figure}

Since {the} small-excursion approximation, in-spite of providing simple and useful expressions \eqref{eq:u_aCBIID}--\eqref{eq:w_aCBIID}, does not provide highly  accurate predictions for a  portion of the parameter space, we devise an alternative approximation technique.  Realizing that a particle located at the free surface {must remain there forever, it} must satisfy $-a_s\!\leq\!z\!\leq\!a_s$. Hence we assume $z=z_0$ (where $z_0$ is particle's initial $z$-position at the free surface) in the eigenfunctions of \eqref{eq:pathline_x_case2}--\eqref{eq:pathline_z_case2}: 

\begin{subequations}
\begin{align}
\frac{d x}{d t} &= k_b \underbrace{\bigg[{A_{s}} \dfrac{\cosh K_b(z_0+H)}{\cosh(K_b H)}+ B_{s}\dfrac{\sinh(K_b z_0)}{\cosh(K_b H)}\bigg]}_{{\mathbb{P}}} \cos {\theta_b} \label{eq:dx/dt},\\
\frac{d y}{d t} &= V_0 + l_b \underbrace{\bigg[{A_{s}} \dfrac{\cosh K_b(z_0+H)}{\cosh(K_b H)}+ B_{s}\dfrac{\sinh(K_b z_0)}{\cosh(K_b H)}\bigg]}_{{\mathbb{P}}} \cos{\theta_b} \label{eq:dy/dt},\\
\frac{d z}{d t} &= K_b \underbrace{\bigg[{A_{s}} \dfrac{\sinh K_b(z_0+H)}{\cosh(K_b H)}+ {B_{s}} \dfrac{\cosh(K_b z_0)}{\cosh(K_b H)}\bigg]}_{{\mathbb{Q}}} \sin{\theta_b} \label{eq:dz/dt}. 
\end{align}
\end{subequations}
We refer to these pathline equations as the `$z$-bounded approximation'. Note that the $z$-bounded approximation circumvents the need of small-excursion assumption, and hence Galilean transformation. Figure \ref{fig:caseII}\,({\it a--d\/}) reveals that the $z$-bounded approximation is highly accurate, and indistinguishable from the exact solution. This is also the case for the deep-water regime as well. Figure \ref{fig:caseII}\,({\it e\/}) shows the particle trajectory in the 3D space. The particle is always located on the free surface, and although is primarily advected along $y$, it does undergo a small drift along $-x$. The undulations on the free surface are the surface impressions of the  {wavy} bottom.

\vspace{5mm}

\paragraph{4.1.2.1. \it{Time period and drift calculations}}
\vspace{4mm}

Like the small-excursion approximation, the $z$-bounded approximation has the advantage of providing a simple expression for the drift velocity, which is otherwise difficult to obtain from the exact equations \eqref{eq:pathline_x_case2}--\eqref{eq:pathline_z_case2}. Equations \eqref{eq:dx/dt} and \eqref{eq:dy/dt} can be combined into a single equation
\begin{equation}
\frac{d \theta_b}{d t}= V_0 l_b+ K_b^2 {{\mathbb{P}}} \cos{\theta_b}, \label{eq:dtheta_b/dt}    
\end{equation}
which is the key to finding the time period, $T_{CBIID}$:
\begin{equation}
    T_{CBIID}=\frac{2\pi}{\sqrt{(V_0 l_b)^2-(K_b^2 {\mathbb{P}})^2}} \approx  T_{aCBIID} \bigg[ 1+ \frac{1}{2}\bigg(\frac{K_b^2 {\mathbb{P}}}{ V_0 l_b}\bigg)^2 \bigg].  \label{eq:Tp}
\end{equation}
Here, $T_{CBIID}$ is the time taken by a particle to complete $2\pi$ phase of its trajectory. In \eqref{eq:Tp}, $K_b^2{\mathbb{P}}\ll V_0l_b$ (evident from \eqref{eq:dy/dt} and \eqref{eq:dtheta_b/dt}), which reveals that $T_{CBIID}$ is slightly longer than the time period $T_{aCBIID}=2\pi/V_0 l_b$ predicted from the small-excursion approximation. To calculate the CBIID velocity, we first divide \eqref{eq:dtheta_b/dt} respectively with \eqref{eq:dx/dt}, \eqref{eq:dy/dt}, and  \eqref{eq:dz/dt}, and evaluate the CBIID displacement  over one time period, $T_{CBIID}$,  by integrating $\theta_b$ from $0$ to $2\pi$. CBIID velocity is obtained after dividing the CBIID displacement by  $T_{CBIID}$:

\begin{subequations}
\begin{align}
    \langle u_{CBIID} \rangle &= - \frac{V_0 k_b l_b}{K_b^2} \Big[1-\sqrt{1-(K_b^2 {\mathbb{P}}/V_0 l_b)^2} \Big] \approx
    \langle u_{aCBIID} \rangle \frac{{\mathbb{P}}^2}{{\mathbb{P}}^2+{\mathbb{Q}}^2},
    \label{eq:u_CBIID}\\
    \langle v_{CBIID} \rangle &=  -\frac{V_0 l_b^2}{K_b^2} \Big[1-\sqrt{1-(K_b^2 {\mathbb{P}}/V_0 l_b)^2} \Big] \approx
     \langle v_{aCBIID} \rangle \frac{{\mathbb{P}}^2}{{\mathbb{P}}^2+{\mathbb{Q}}^2},
     \label{eq:v_CBIID}\\
    \langle w_{CBIID} \rangle &=0, \label{eq:w_CBIID}
\end{align}
\end{subequations}

\begin{figure}
  \centering
  \begin{subfigure}{0.8\textwidth}
  \centering
  \includegraphics[width=\linewidth]{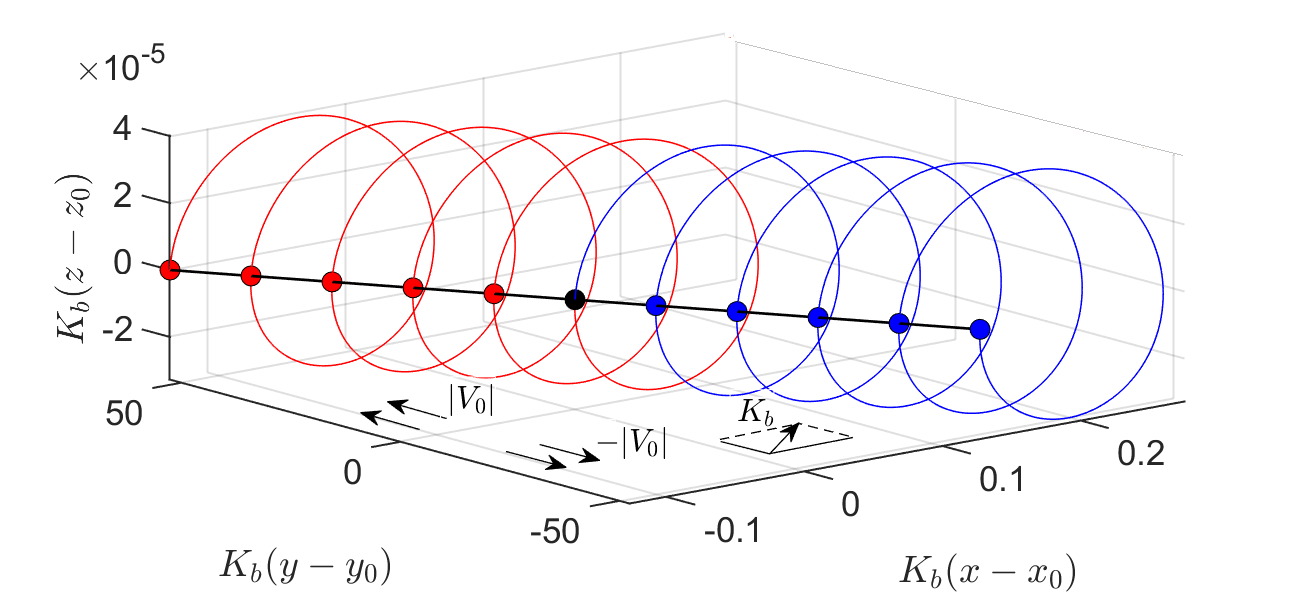}
  \caption{}
  \end{subfigure}
  \hspace{-2mm}
  \begin{subfigure}{0.51\textwidth}
  \centering
  \includegraphics[width=\linewidth]{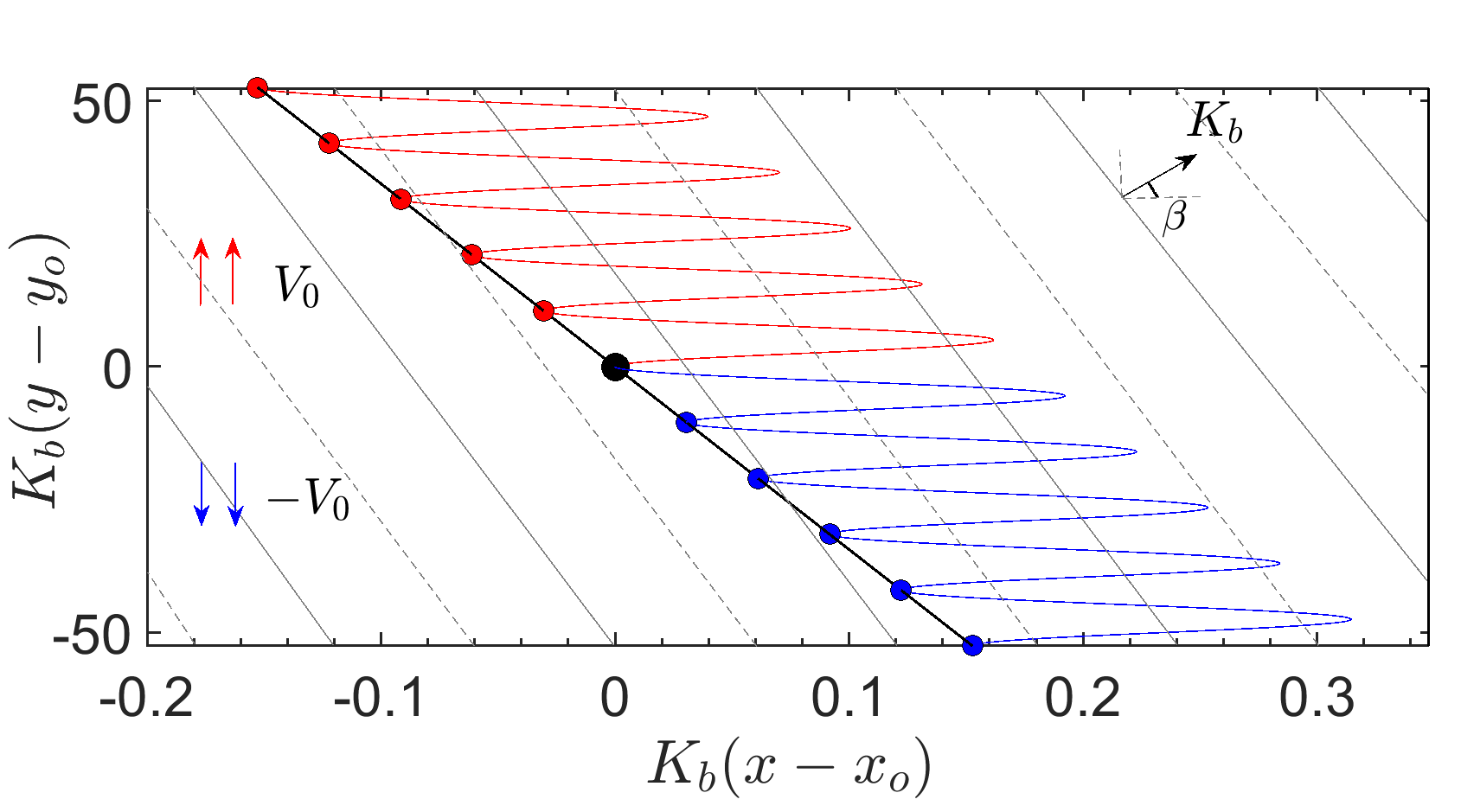}
  \caption{}
  \end{subfigure}
  \begin{subfigure}{0.48\textwidth}
  \centering
  \includegraphics[width=\linewidth]{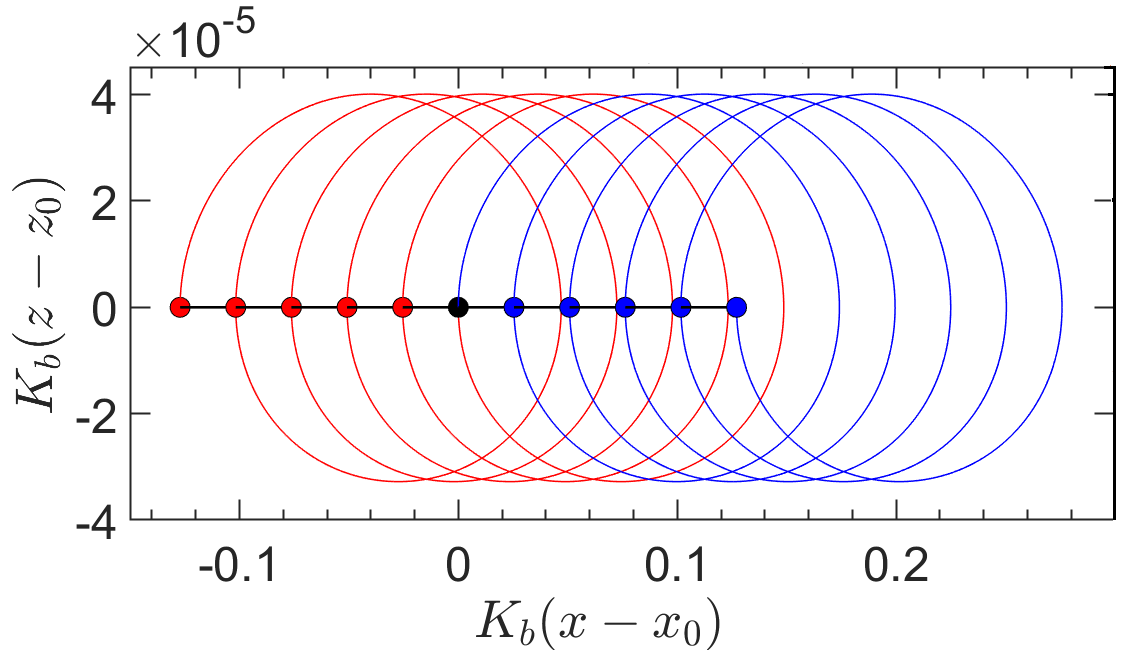} 
  \caption{} 
  \end{subfigure}
  \caption{\footnotesize  Particle trajectory for case-II in the stationary reference frame. Particle trajectory (a) in the  3D space, (b) in the  $x$--$y$ plane, and (c) in the  $ x$--$z$ plane. Red (blue) solid line indicates trajectory when $V_0>0$ ($V_0<0$). The particle's initial position is shown by filled black circle,  while red (blue) circle indicates particle's position after each time period ($T_{CBIID}$) for $V_0>0$ ($V_0<0$). Parameters used: $K_b H (k_b H, l_b H)=0.1 (0.08,0.06)$, $a_b/H=0.1$, $a/H=0$, $Fr=0.1$.}
  \label{fig:caseII_background_current}
\end{figure}
 
\noindent The above equations show that  $\langle \boldsymbol{u}_{CBIID} \rangle$ is not exactly the same as  $\langle \boldsymbol{u}_{aCBIID} \rangle$, and hence provide another evidence for the disparity between the exact solution and the small-excursion approximation in figure \ref{fig:caseII}\,({\it a--d\/}). Since shallow-water/{long--bottom-undulation} limit, $k_bH\ll 1$ and $l_bH \ll 1$, and relatively high (but still a small quantity) $a_b/H$ produces the maximum disparity (as shown in figure \ref{fig:caseII}\,{\it d\/}), the order of magnitude of this discrepancy needs to be evaluated.  After a detailed but straight-forward algebra, we obtain
\begin{equation}
    T^{shallow}_{CBIID} \approx  T_{aCBIID}[ 1+ \mathrm{O}((a_b/H)^2)],\,\, \mathrm{and}\,\, \langle u_{CBIID}^{shallow} \rangle =\langle u_{aCBIID} \rangle [1+ \mathrm{O}(\epsilon_b^2)].
\end{equation}
Hence the relative difference between the two time periods scale with  $(a_b/H)^2$, confirming the discrepancies and the trend observed in figures  \ref{fig:caseII}\,({\it c\/}) and \ref{fig:caseII}\,({\it d\/}).
Since $\epsilon_b=(k_bH)(a_b/H)\ll a_b/H$ (in the shallow-water limit), the error in evaluating $\langle u_{aCBIID} \rangle$ is far less in comparison to that of $T_{aCBIID}$. 

\vspace{3mm}
\subsection{Parametric analysis}

\vspace{3mm}
\subsubsection{Effect of the alongshore current, $V_0$}

Equation \eqref{eq:u_CBIID} reveals that the magnitude of 
$\langle u_{CBIID} \rangle$ is directly proportional to $V_0$, hence stronger alongshore current will produce stronger  CBIID velocity in the cross-shelf direction. The sign of $V_0$ determines whether  particles will drift along $+x$ or $-x$. When $V_0>0$ ($V_0<0$), {$\langle u_{CBIID} \rangle$ is negative (positive), i.e. particles will move in the $-x$ ($+x$) direction,
see figure \ref{fig:caseII_background_current}\,({\it a--c\/}).}

\subsubsection{Effect of the bottom topography wave-vector, $\vect{K_b}$}

Equation \eqref{eq:u_CBIID}, 
\begin{equation}
 \langle u_{CBIID} \rangle = - \frac{V_0}{2}  \sin (2\beta) \Big[1-\sqrt{1-(K_b^2 {\mathbb{P}}/V_0 l_b)^2} \Big],  
   \label{eq:422_rewrite}
\end{equation}
re-written here for convenience, shows the dependence of the cross-shelf drift velocity on $\beta\equiv \tan^{-1}(l_b/k_b)$. Particle trajectories are plotted in  figure \ref{fig:caseII_background_current}\,({\it a\/}) for $\beta \in [0,\pi/2]$ and $V_0$ both positive and negative, thereby allowing the angle between $\vect{K_b}$ and $V_0$ to span between $0$ and $\pi$.

The magnitude of the wave-vector, i.e. $K_b$, can influence $\langle u_{CBIID} \rangle$ via the term `[$\ldots$]' in \eqref{eq:422_rewrite}. This bracketed term, for shallow and deep water limits, are respectively as follows:
$$ [\ldots]_{shallow} \approx \dfrac{1}{2} \bigg( \dfrac{a_b/H}{Fr^2 \sin^2\beta - 1} \bigg)^2,  $$
$$ \, [\ldots]_{deep} \approx \dfrac{1}{2} \big( a_b K_b \ee^{-K_bH} \big)^2. $$

\noindent Hence  the magnitude of $K_b$ does not affect  the cross-shelf drift velocity in the shallow-water limit, but does affect in the deep-water limit.  
{This behaviour is confirmed in figure  \ref{fig:parameters_comparison_contours}\,({\it a\/}), where $\langle u_{CBIID} \rangle$ contours are plotted for a particle at the free surface.
Figure  \ref{fig:parameters_comparison_contours}\,({\it a\/}) clearly reveals that for a given $K_bH$ value, the maximum drift always occur at $\beta \approx \pm \pi/4$. This is not surprising because the coefficient $\sin(2\beta)$ in  \eqref{eq:422_rewrite} attains  its maximum value when $\beta=\pm\pi/4$ (the term in square brackets in \eqref{eq:422_rewrite}  has a very weak dependence on $\beta$, thereby causing a slight deviation from $\pm\pi/4$). Additionally, figure  \ref{fig:parameters_comparison_contours}\,({\it a\/}) also reveals that for a given $\beta$, the longer the bottom undulations, the higher is the drift. In summary, long bottom undulations with $\beta \approx \pm\pi/4$ causes maximum  $\langle u_{CBIID} \rangle$.}

{ While investigating interactions between different kinds of bottom topography with a shear current,  \cite{akselsen2019sheared} investigated a case with sinusoidal bottom having $\beta=\pi/4$ (this was the only  $\beta$ value the authors have considered). The authors observed  helical curving and spanwise migration of streamlines due to 3D bed current vorticity interactions. The authors also point out that this effect disappears only if the current is shear free, i.e., uniform. Since our analysis is  based only on uniform background current (and all perturbations are irrotational), CBIID is quite different from the mechanism observed in \cite{akselsen2019sheared}.}

\vspace{4mm}
\subsubsection{Effect of bottom  {topography} amplitude, $a_b$}

\begin{figure}
  \centering
  \hspace{-3mm}
  \begin{subfigure}{0.45\textwidth}
  \centering
  \includegraphics[width=\linewidth]{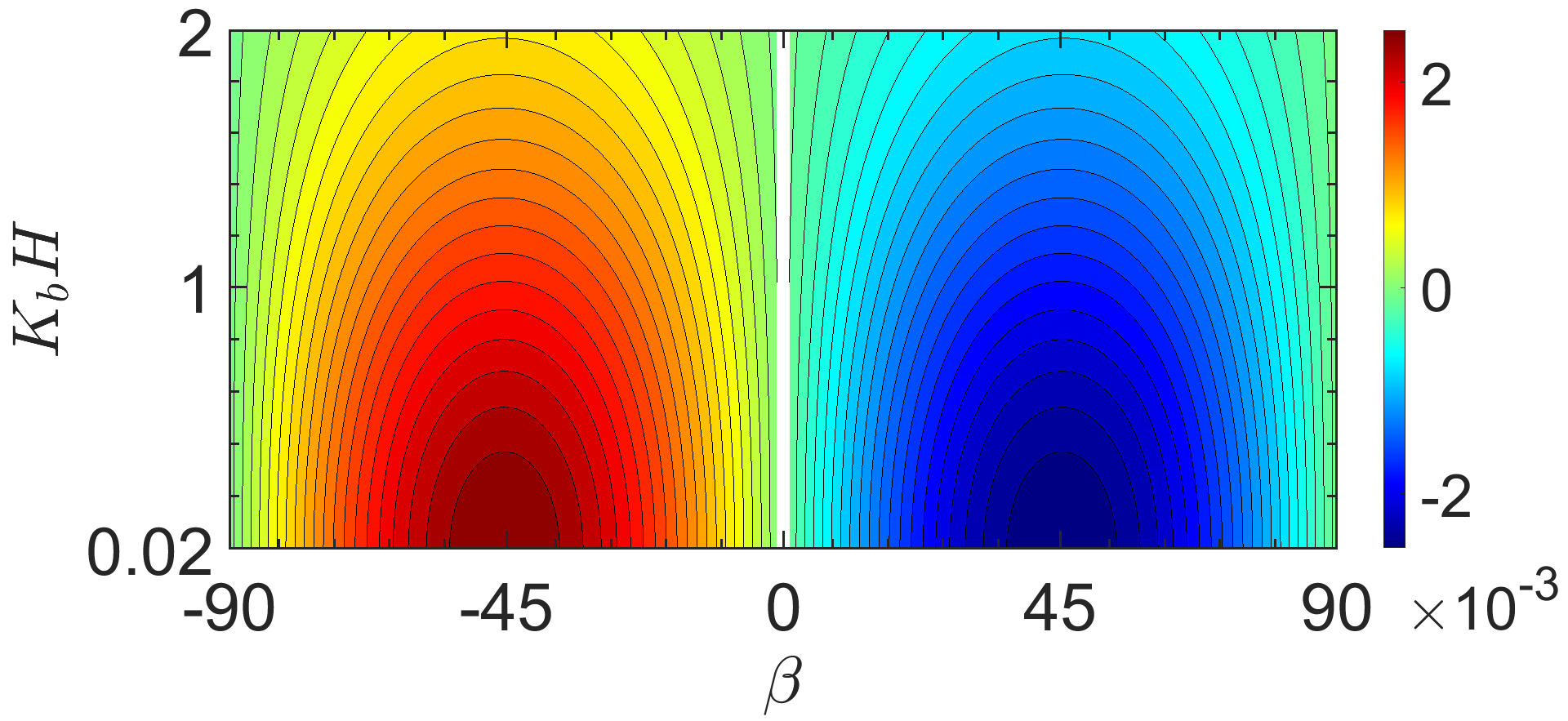}
  \caption{} 
  \end{subfigure}
  \hspace{-2mm}
  \begin{subfigure}{0.45\textwidth}
  \centering
  \includegraphics[width=\linewidth]{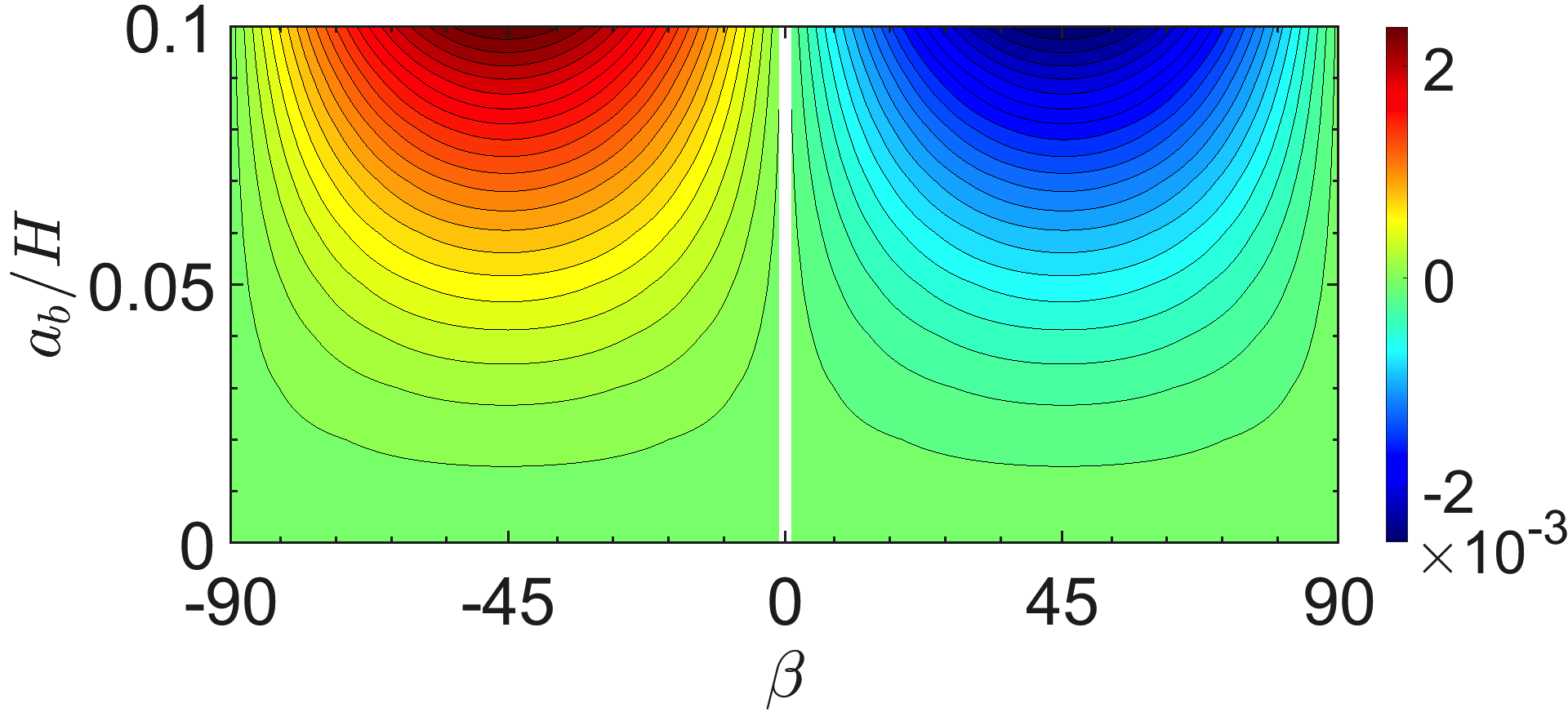}
  \caption{} 
  \end{subfigure}
  \hspace{-2mm}
  \begin{subfigure}{0.45\textwidth}
  \centering
  \includegraphics[width=\linewidth]{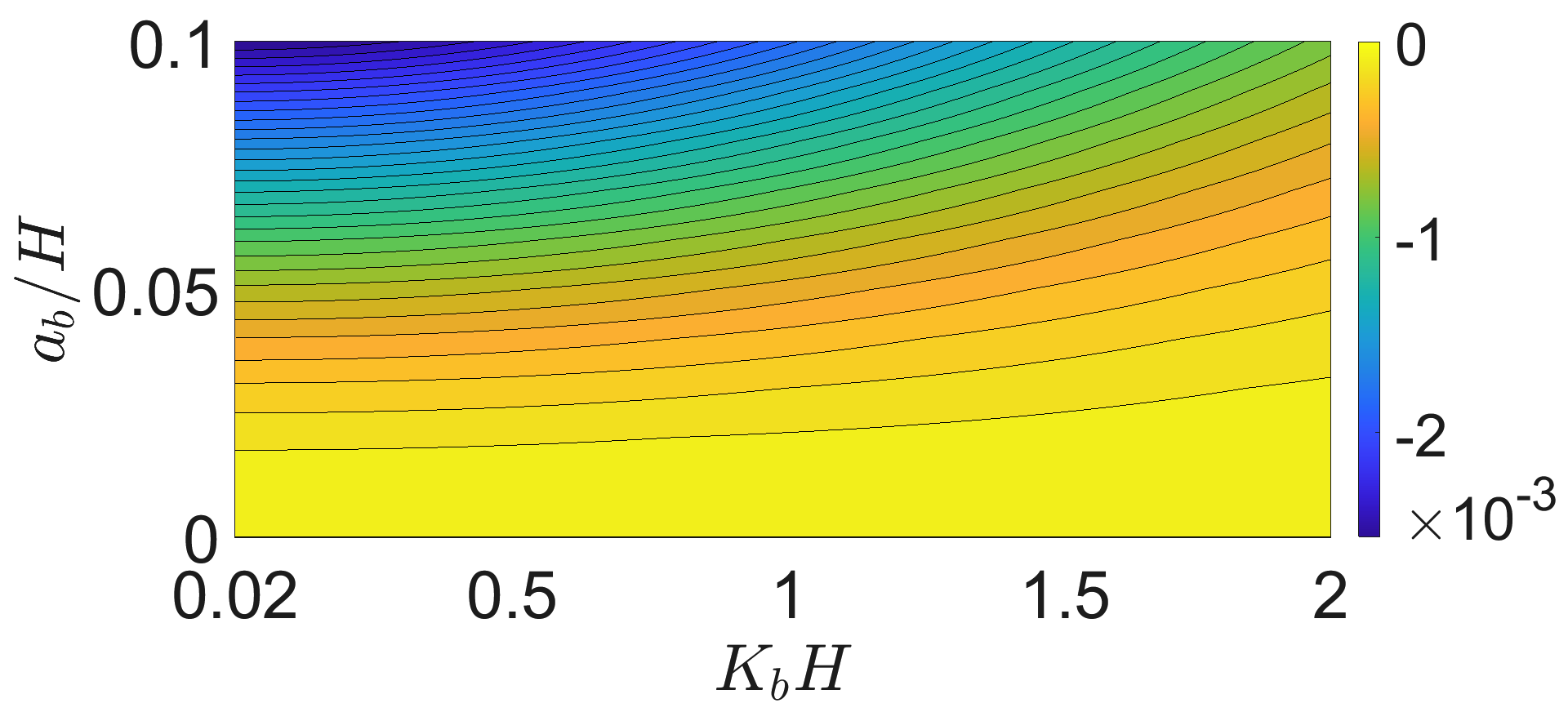}
  \caption{} 
  \end{subfigure}
  \begin{subfigure}{0.45\textwidth}
  \centering
  \includegraphics[width=\linewidth]{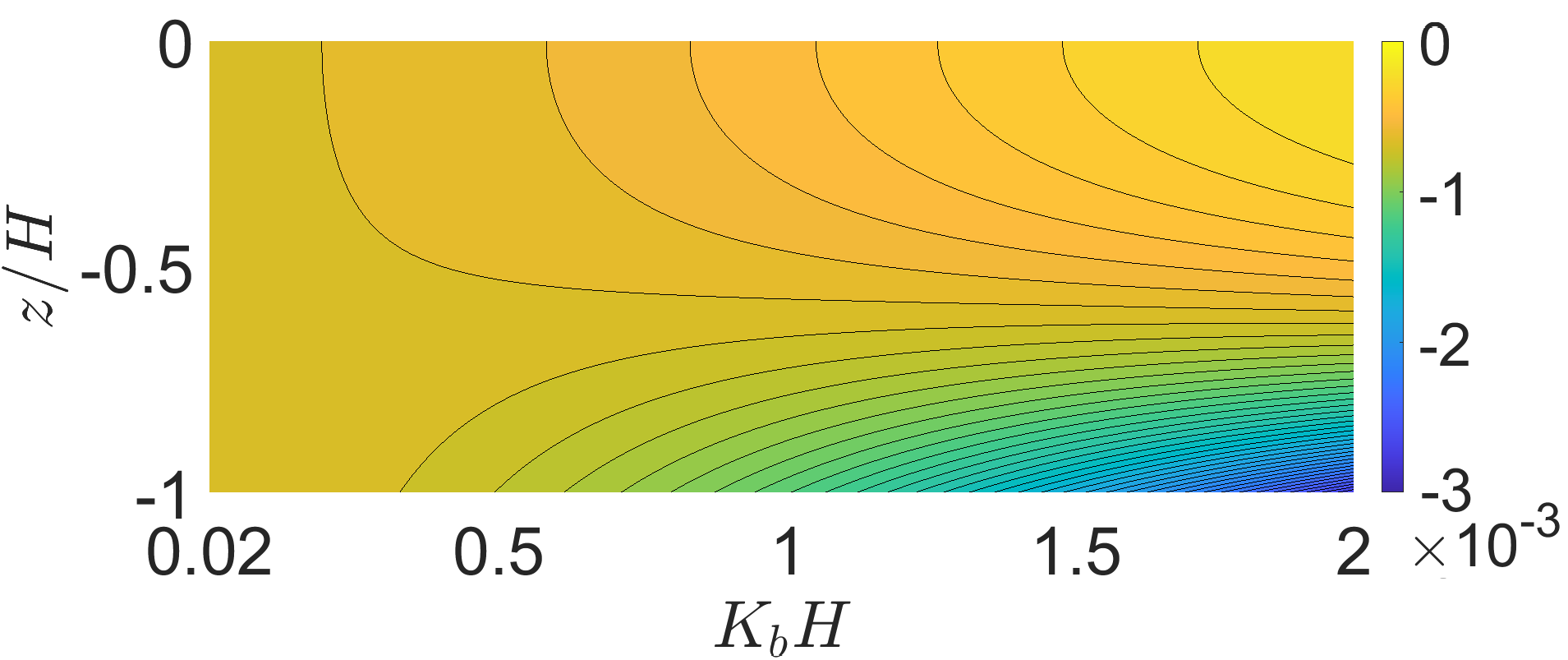}
  \caption{}
  \end{subfigure}
  \caption{\footnotesize Contour plots of  $\langle u_{CBIID} \rangle/V_0$ in the  (a) $\beta$--$K_b H$ plane for $a_b=0.1H$ at $z_0=0$, (b) $\beta$--$a_b/H$ plane for $K_bH=0.2$ at $z_0=0$,
  (c) $K_b H$--$a_b/H$ plane for $\beta=45\degree$ at $z_0=0$, and (d) $K_b H$--$z/H$ plane for $\beta=45\degree$ and $a_b=0.05H$. For all plots,  $Fr=0.1$ ($V_0>0$).} 
  \label{fig:parameters_comparison_contours}
\end{figure}

Equation \eqref{eq:u_aCBIID} reveals that CBIID velocity in the cross-shelf direction is proportional to $a_b^2$ (since  ${\mathbb{P}}^2$ and ${\mathbb{Q}}^2$ are both proportional to $a_b^2$), highlighting the crucial role played by bottom {topography's} amplitude in cross-shelf transport. {The variation of $\langle u_{CBIID} \rangle$ with the amplitude of bottom undulations is shown in figure \ref{fig:parameters_comparison_contours}\,({\it b\/}--{\it c\/}). The contour plots do reveal that higher $a_b/H$ values lead to higher CBIID, and that maximum drift corresponds to long-bottom-undulation limit with $\beta \approx \pm \pi/4$.    }

\vspace{3mm}
\subsubsection{Effect of particle's initial $z$-location, $z_0$}

The fact that $|u^{(1)}|$ is maximum at the bottom and decays with elevation, shown in figure \ref{fig:contour_plots}\,({\it b\/}),  indicates that CBIID
will have a similar variation. {  Contour plot of  $\langle u_{CBIID} \rangle$, shown in figure  \ref{fig:parameters_comparison_contours}\,({\it d\/}), reveals that this is indeed the case. For intermediate ($K_bH\approx 1$) or deep ($K_bH\gg 1$)  water situations,  particles initially located at greater depths, e.g.\, submerged particles like sediments, will experience higher drift velocities than floating particles. For shallow water situation ($K_bH\ll 1$), variations with depth is non-existent.   Figure \ref{fig:parameters_comparison_contours}\,({\it d\/}) also reveals that near the bottom, $\langle u_{CBIID} \rangle$ increases with an increase in $K_bH$, while near the free surface, $\langle u_{CBIID} \rangle$ decreases when $K_bH$ increases. This behavior of CBIID is just the reverse of Stokes drift. 
}


\section{Case-III: Wave steepness and  {wavy} seabed's steepness of the same order of magnitude [$\mathrm{O}(\epsilon_b) \sim \mathrm{O}(\epsilon) { \ll 1}$]}
\label{sec:case3}

Here we consider a situation shown in figure \ref{fig:Schmatic_diagram}\,({\it b\/}) where the wave steepness ($\epsilon$) is of the same order as the {wavy} seabed steepness ($\epsilon_b$); i.e. $\mathrm{O}(\epsilon_b) \sim \mathrm{O}(\epsilon) {\ll 1}$. To study the linear interaction of the uniform alongshore current with the  {wavy} bottom topography and surface waves,  
we substitute the perturbation series of $\phi$ and $\eta$  from \eqref{eq: phi_full}--\eqref{eq: eta_full} into GLE and BCs, given in \eqref{eq:ImC}, \eqref{eq:KBC}--\eqref{eq:DBC}. At $\mathrm{O}(\epsilon)$ or $\mathrm{O}(\epsilon_b)$, we obtain:
\begin{subequations}
\begin{align}
& \phi_{,xx}^{(1)}+\phi_{,yy}^{(1)}+\phi_{,zz}^{(1)}=0 \qquad  & -H < z < 0,\\
& \phi_{,z}^{(1)}=V_0 \eta_{b,y} \qquad &\mathrm{at} \, z=-H,\\
&  \eta_{,t}^{(1)}+V_0\eta_{,y}^{(1)}-\phi_{,z}^{(1)}=0  \qquad &\mathrm{at} \, z=0, \\
&  \phi_{,t}^{(1)}+V_0 \phi_{,y}^{(1)}+g\eta^{(1)}=0 \qquad &\mathrm{at} \, z=0.
\end{align}
\end{subequations}
Next, we assume the {wavy} bottom topography to be the same as that in case-II (given in \eqref{eq:etab_def}), and the surface wave profile to the same as that considered in case-I (given in \eqref{eq:eta_case1}).  Hence the combined (i.e. unsteady $+$ steady) solutions of the surface elevation and velocity potential are respectively as follows: 
\begin{subequations}
\begin{align}
\eta^{(1)}=\eta_{u}^{(1)}+\eta_{s}^{(1)}=&\,  a \cos{\theta}+ a_{s} \cos{\theta_b},\label{eq:eta_combined}\\
\phi^{(1)}=\phi_{u}^{(1)}+\phi_{s}^{(1)}=&\, \frac{a \overline{\omega}}{K} \frac{\cosh K(z+H)}{\sinh (KH)} \sin{\theta}+
\bigg[{A_{s}} \frac{\cosh K_b(z+H)}{\cosh (K_b H)}  + {B_{s}} \frac{\sinh (K_b z)}{\cosh (K_b H)} \bigg] \sin{\theta_b}. \label{eq:phi_combined}
\end{align}
\end{subequations}
Figures \ref{fig:contour_plots}\,({\it c\/}) and \ref{fig:contour_plots}\,({\it f\/}) respectively show contour plots of cross-shelf velocity, $u^{(1)}$($=\phi_{,x}^{(1)}$, where $\phi^{(1)}$ is from \eqref{eq:phi_combined}) for intermediate and shallow depths.
While  figure \ref{fig:contour_plots}\,({\it a\/}) reveals that $|u^{(1)}|$ decreases with depth, and figure \ref{fig:contour_plots}\,({\it b\/}) shows exactly the reverse, their combination, figure \ref{fig:contour_plots}\,({\it c\/}), show a non-monotonic variation in $|u^{(1)}|$ -- surface waves dominating in the upper layer while current-bathymetry interactions dominating the bottom layer.

\subsection{Pathline equations}

The pathline equations can be obtained by substituting $\boldsymbol{u^{(1)}}(=\nabla \phi^{(1)}$)  into \eqref{eq:pathline}, which are as follows:
\begin{subequations}
\begin{align}
    \frac{d x}{d t}=&\frac{a \overline{\omega} k}{K} \frac{\cosh K(z+H)}{\sinh (KH)} \cos{\theta}+ k_b \bigg[{A_{s}} \frac{\cosh K_b(z+H)}{\cosh (K_b H)}  + {B_{s}} \frac{\sinh (K_b z)}{\cosh (K_b H)} \bigg] \cos{\theta_b}, \label{eq:pathline_case3_1}\\
    \frac{d y}{d t}=&V_0+\frac{a \overline{\omega} l}{K} \frac{\cosh K(z+H)}{\sinh (K H)} \cos{\theta}+  l_b \bigg[{A_{s}} \frac{\cosh K_b(z+H)}{\cosh (K_b H)}  + {B_{s}} \frac{\sinh (K_b z)}{\cosh (K_b H) } \bigg] \cos{\theta_b}, \label{eq:pathline_case3_2}\\
    \frac{d z}{d t}=&{a \overline{\omega}} \frac{\sinh K(z+H)}{\sinh (K H)} \sin{\theta}+  K_b \bigg[{A_{s}} \frac{\sinh K_b(z+H)}{\cosh (K_b H)}  + {B_{s}} \frac{\cosh (K_b z)}{\cosh (K_b H)} \bigg] \sin{\theta_b}.
    \label{eq:pathline_case3_3}
\end{align}
\label{eq:pathline_case3}
\end{subequations}

\begin{figure}
  \centering
  \begin{subfigure}{0.75\textwidth}
  \centering
  \includegraphics[width=\linewidth]{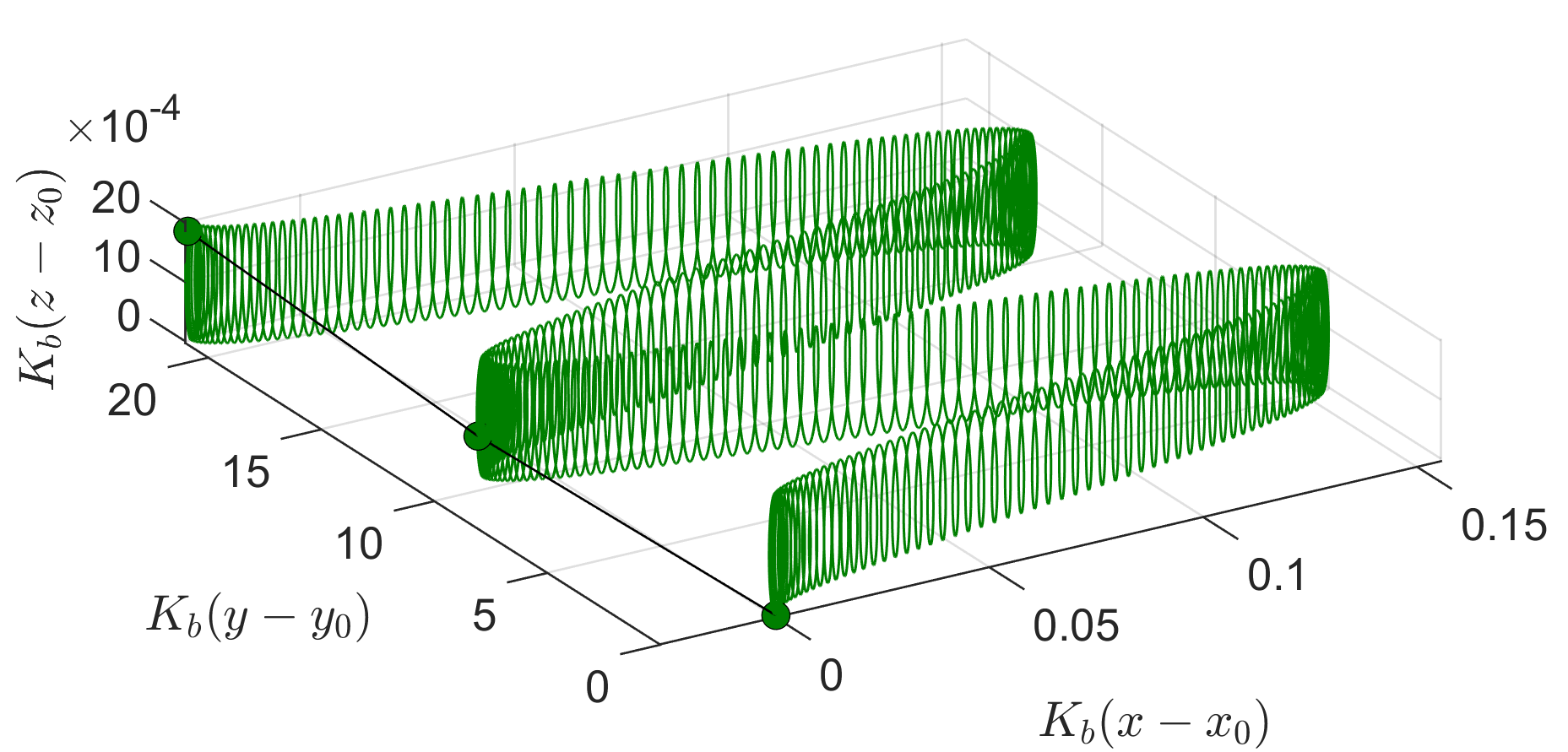}
  \caption{} \label{fig:CaseIII_3D}
  \end{subfigure}
  \begin{subfigure}{0.8\textwidth}
  \centering
  \includegraphics[width=\textwidth]{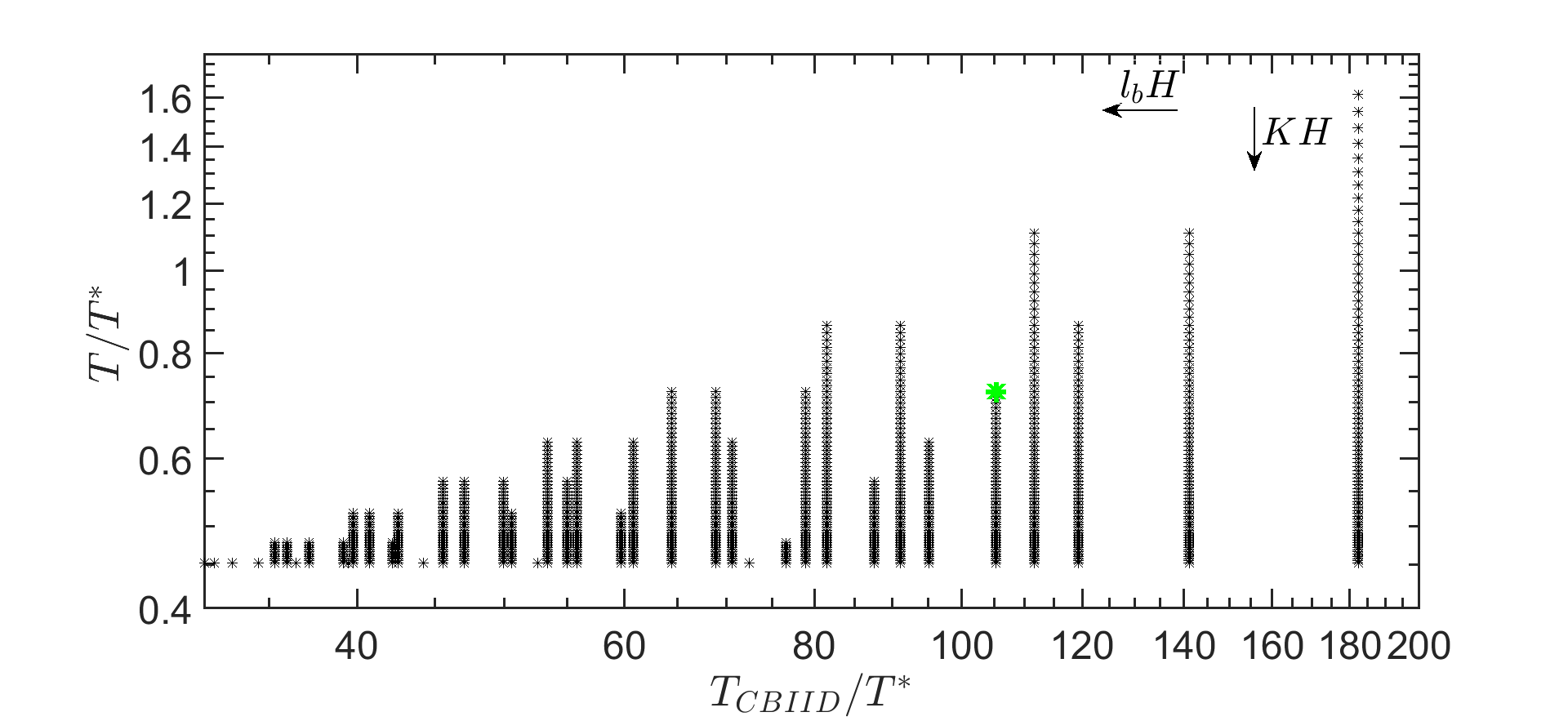}
  \caption{} 
  \end{subfigure}
  \caption{\footnotesize Case-III with $a_b=0.1H$ and $Fr=0.1 \, (V_0>0)$. (a)   
  Particle trajectory in the non-dimensional 3D space, {denoted by the solid green curve}, is plotted for two time period{(s)}. 
  Filled green circles denote positions after each $T_{CBIID}$, and are connected by the Lagrangian mean trajectory. Here $a=0.01H$, and $KH (kH,lH)=1 (1,0)$, $K_b H (k_b H, l_b H)=0.1 (0.08, 0.06)$. (b) $T_{CBIID}$ v/s $T$ plot for $a \le 0.01H$, and the following range of wavenumbers: $KH (kH,0)=0.2-2$,  $l_bH=0.02-0.2$. The green asterisk shows the case corresponding to sub-figure (a), and {$T^*=H/V_0$ is the advection time scale.}}   
  \label{fig:CaseIII_particle_motion}
\end{figure}

Figure \ref{fig:CaseIII_particle_motion}\,({\it a\/}) shows the trajectory plot obtained by solving the  pathline equations \eqref{eq:pathline_case3_1}--\eqref{eq:pathline_case3_3}. The figure
reveals two spatial scales -- one due to surface waves (fast oscillations) and the other due to CBIID (slow oscillations). To understand the reason behind the occurrence of two spatial scales, we perform a scatter plot of the two temporal scales, $T$ and $T_{CBIID}$, in figure \ref{fig:CaseIII_particle_motion}\,({\it b\/}). We scanned a range of parameters (see caption of figure \ref{fig:CaseIII_particle_motion}\,{\it b\/}) for which $a/H \lesssim \mathrm{O}(\epsilon)$ is satisfied, and we found $T$ and $T_{CBIID}$ to be order separated. While $T$ is in the order of a few seconds (which is well known),  $T_{CBIID}$ is typically of the order of a few minutes for realistic nearshore parameters. The drift velocities, $u_{SD}$ and $u_{CBIID}$, are found to have similar magnitudes, hence the spatial scale separation in  figure \ref{fig:CaseIII_particle_motion}\,({\it a\/}) is exclusively due to the two time scales of the problem.  

\vspace{2mm}
\subsubsection{Combined drift: small-excursion approximation}

\S \ref{subsec:4.1.1} has already revealed that {the} small-excursion approximation might not provide highly accurate estimates, however in case-III, it is probably the only choice. While applying {the} small-excursion approximation about an initial particle location $\boldsymbol{X_0}=(x_0, y_0, z_0)$, we first need to implement a Galilean transformation: $(X,Y,Z)=(x,y-V_0t,z)$ to the pathline equation \eqref{eq:pathline_case3_1}--\eqref{eq:pathline_case3_3}, and Taylor expand the pathline equations about the initial position. The particle motion depends on the combined effect of surface waves and current--bathymetry interaction; therefore, we refer to this drift as a combined drift (CD). The \emph{approximate}-CD (aCD) velocity can be defined as, 

\begin{equation}
\boldsymbol{\langle u_{aCD} \rangle}
= \langle \boldsymbol{(X-X_0)} \cdot \boldsymbol{\nabla u^{(1)}}|_{\boldsymbol{X=X_0}} \rangle,
\label{eq:combined_drift_formula}
\end{equation}
where $$\boldsymbol{X=X_u+X_s},$$  
$$\boldsymbol{\nabla u^{(1)}}=\left. \begin{bmatrix}
u_{u,X}^{(1)} + u_{s,X}^{(1)} & v_{u,X}^{(1)} + v_{s,X}^{(1)} & w_{u,X}^{(1)} + w_{s,X}^{(1)}\\ \\ 
u_{u,Y}^{(1)} + u_{s,Y}^{(1)} & v_{u,Y}^{(1)} + v_{s,Y}^{(1)} & w_{u,Y}^{(1)} + w_{s,Y}^{(1)}\\ \\ 
u_{u,Z}^{(1)} + u_{s,Z}^{(1)} & v_{u,Z}^{(1)} + v_{s,Z}^{(1)} & w_{u,Z}^{(1)} + w_{s,Z}^{(1)}
\end{bmatrix}  \right. ,$$
and $\langle \ldots \rangle$ denotes averaging over one time period, $T_{aCD}$, 
and each term of \eqref{eq:combined_drift_formula} is discussed in  detail in appendix \ref{appendixsec:small_excursion}.

While $T_{aCD}=\mathrm{LCM}(\overline{T},T_{aCBIID})$ is the formal method for evaluation, its implementation is challenging in practice since $\overline{T}$ and $T_{aCBIID}$ are, in general, not integers (in fact, they are typically irrational numbers). The practical approach would be to approximate $\overline{T}/T_{aCBIID}$ by its nearest rational number, $n/m$ (i.e. $m,n \in \mathbb{Z}^+$). This leads to  $T_{aCD} \approx n T_{aCBIID} \approx m \overline{T}$, which is crucial for performing the averaging. Finally this yields (see appendix \ref{appendixsec:small_excursion})
\begin{equation}
\boldsymbol{\langle u_{aCD} \rangle} = \boldsymbol{\langle u_{SD} \rangle}+\boldsymbol{\langle u_{aCBIID} \rangle}.
\label{app eq:app_CD2}
\end{equation}

{In a realistic nearshore environment, we need to account for the Eulerian return flow for calculating the net Lagrangian drift velocity in the cross-shelf direction:  }
\begin{equation}
    {U_L = U_E + \underbrace{\langle u_{SD} \rangle +  \langle u_{aCBIID} \rangle}_{\langle u_{aCD} \rangle},}
    \label{eq:UL_caseIII}
\end{equation}
{where $U_E= U_{E(SD)}+U_{{E(CBIID)}}$ is the net Eulerian return flow.}


\section{{Comparison of the three cases for realistic nearshore parameters}}
\label{sec:casestudy}

\begin{figure}
  \centering
  \begin{subfigure}{0.48\textwidth}
  \centering
  \includegraphics[width=\linewidth]{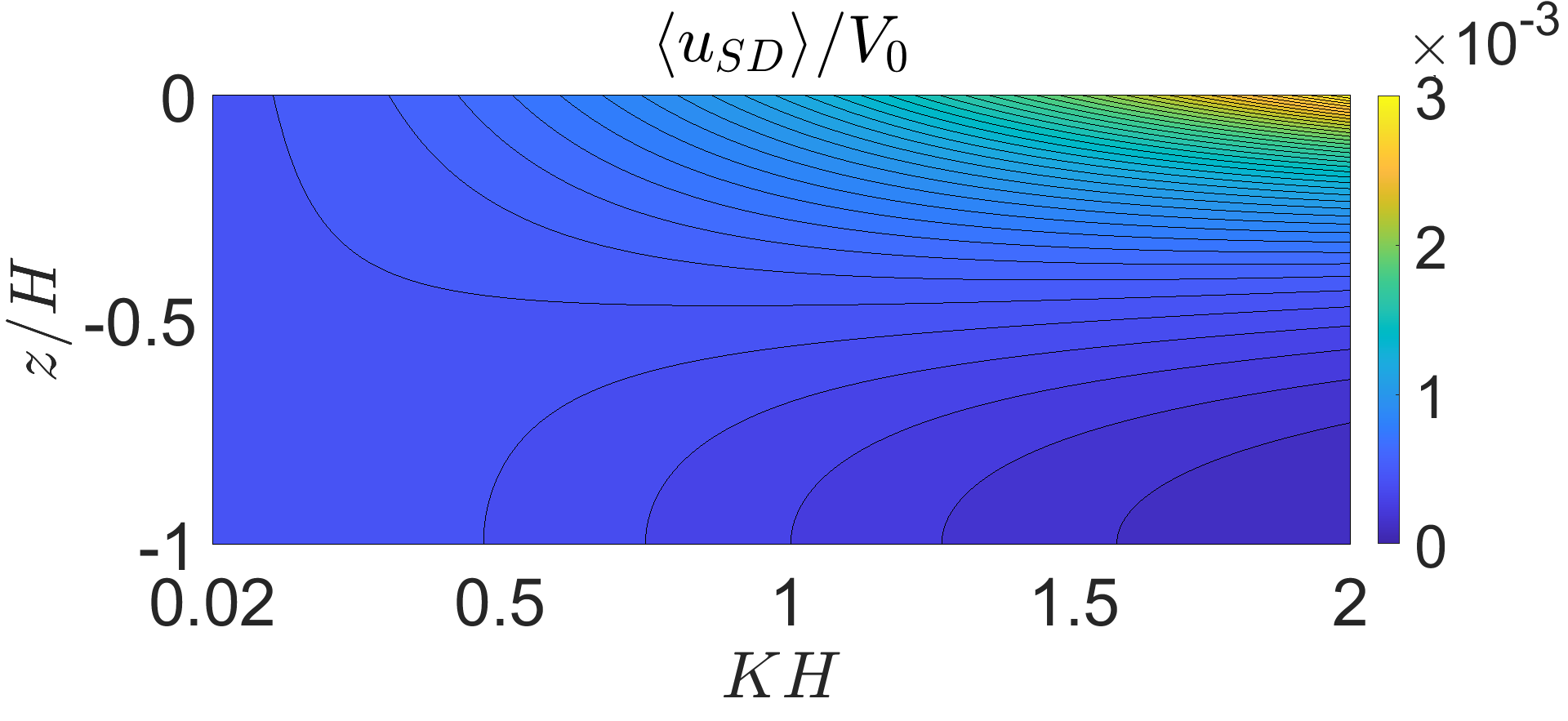}
  \caption{}
  \end{subfigure}
  \begin{subfigure}{0.48\textwidth}
  \centering
  \includegraphics[width=\linewidth]{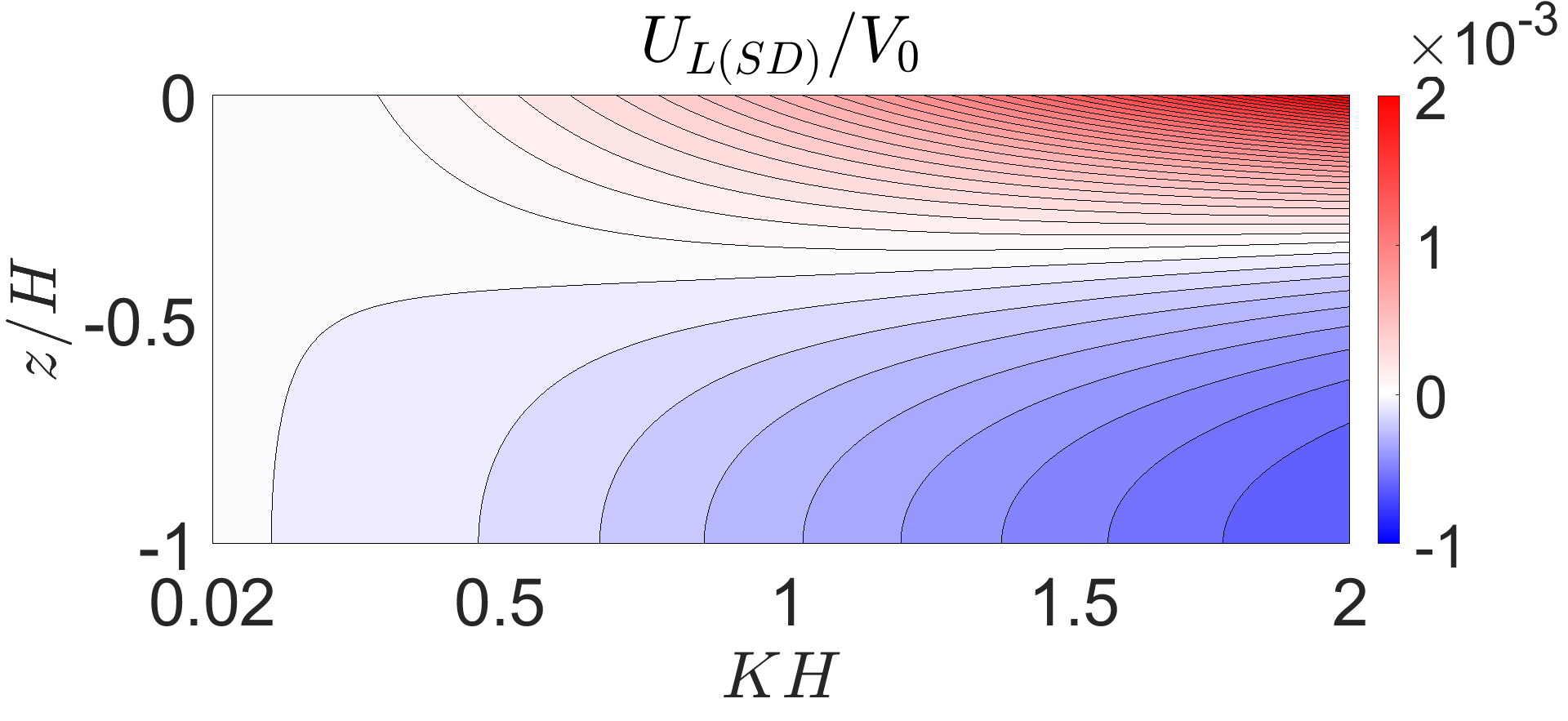}
  \caption{} 
  \end{subfigure}
  \begin{subfigure}{0.48\textwidth}
  \centering
  \includegraphics[width=\linewidth]{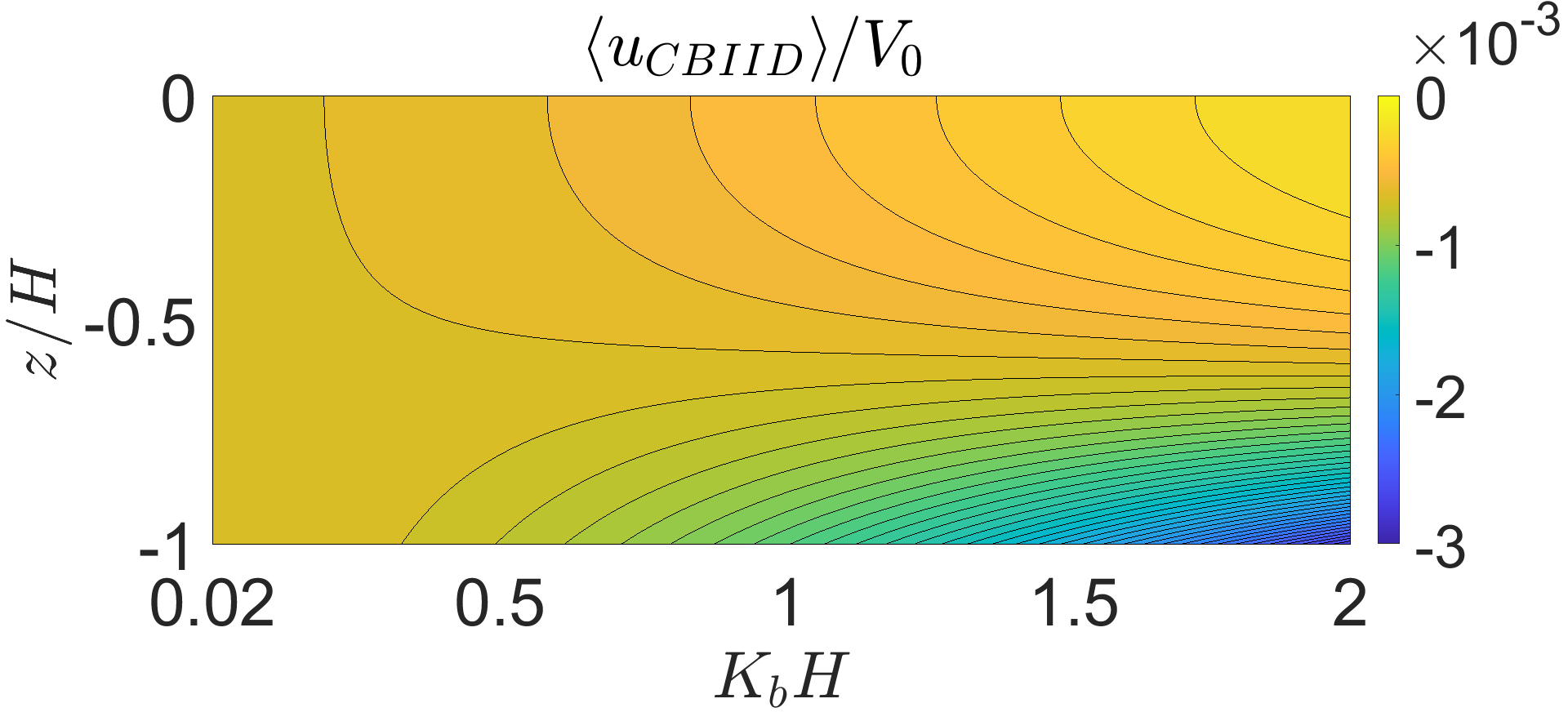}
  \caption{} 
  \end{subfigure}
  \centering
  \begin{subfigure}{0.48\textwidth}
  \centering
  \includegraphics[width=\linewidth]{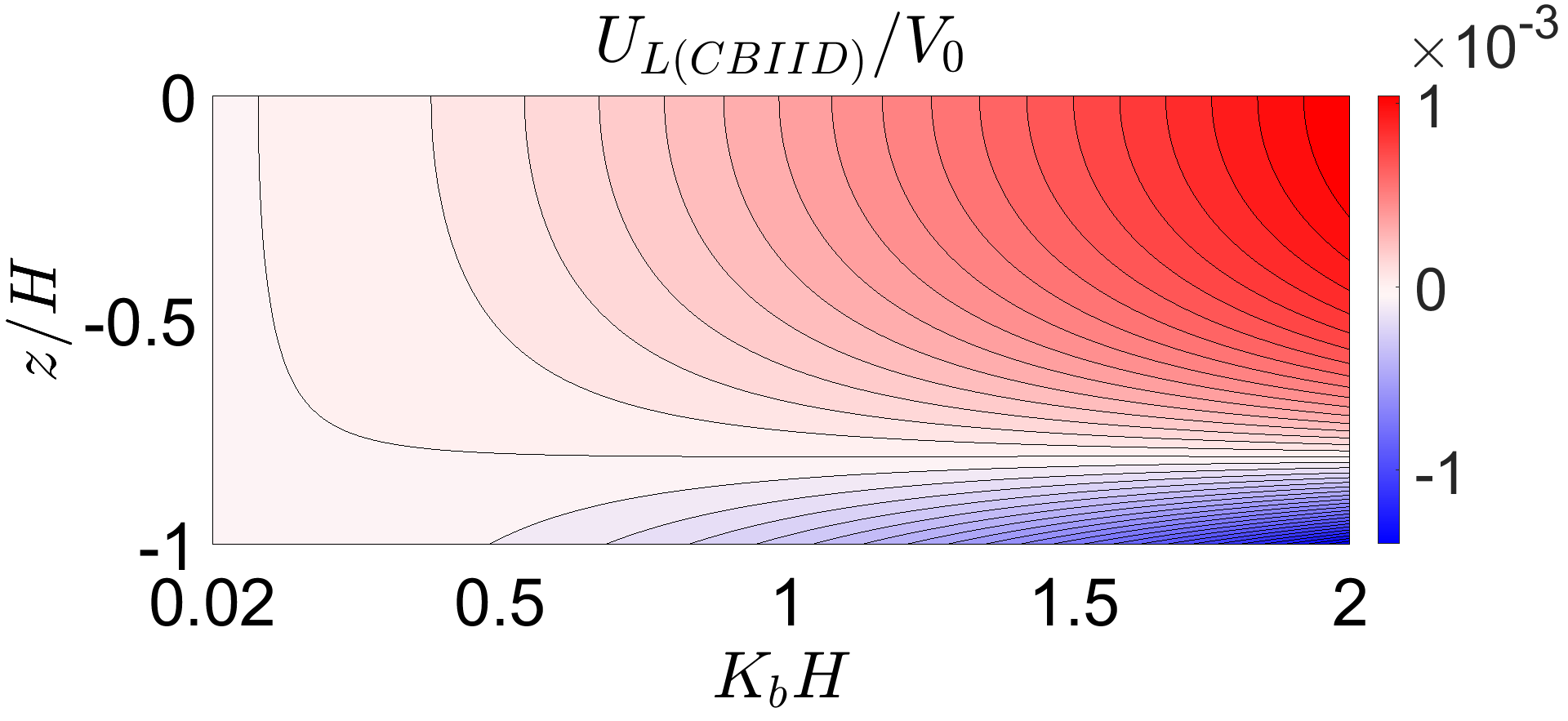}
  \caption{}
  \end{subfigure}
  \begin{subfigure}{0.48\textwidth}
  \centering
  \includegraphics[width=\linewidth]{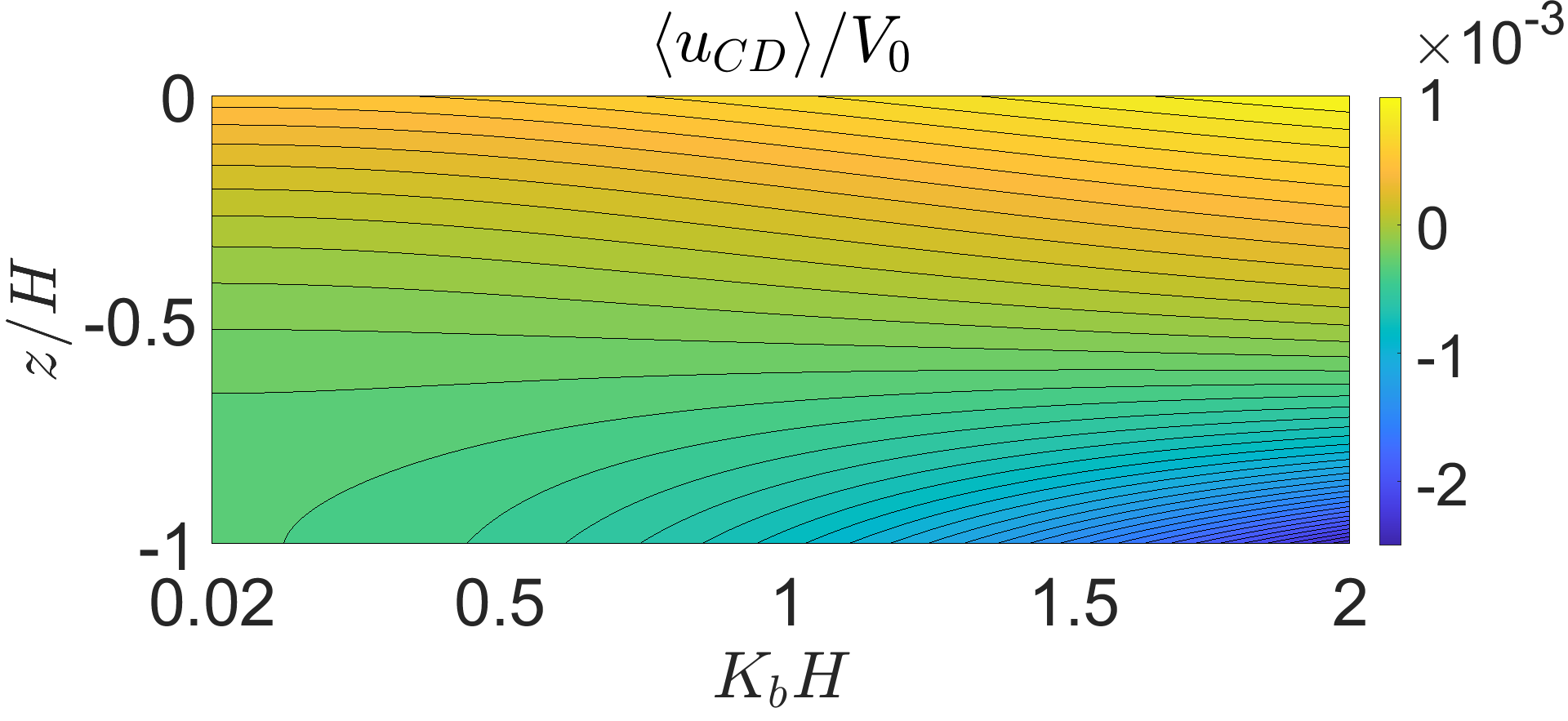}
  \caption{} 
  \end{subfigure}
  \begin{subfigure}{0.48\textwidth}
  \centering
  \includegraphics[width=\linewidth]{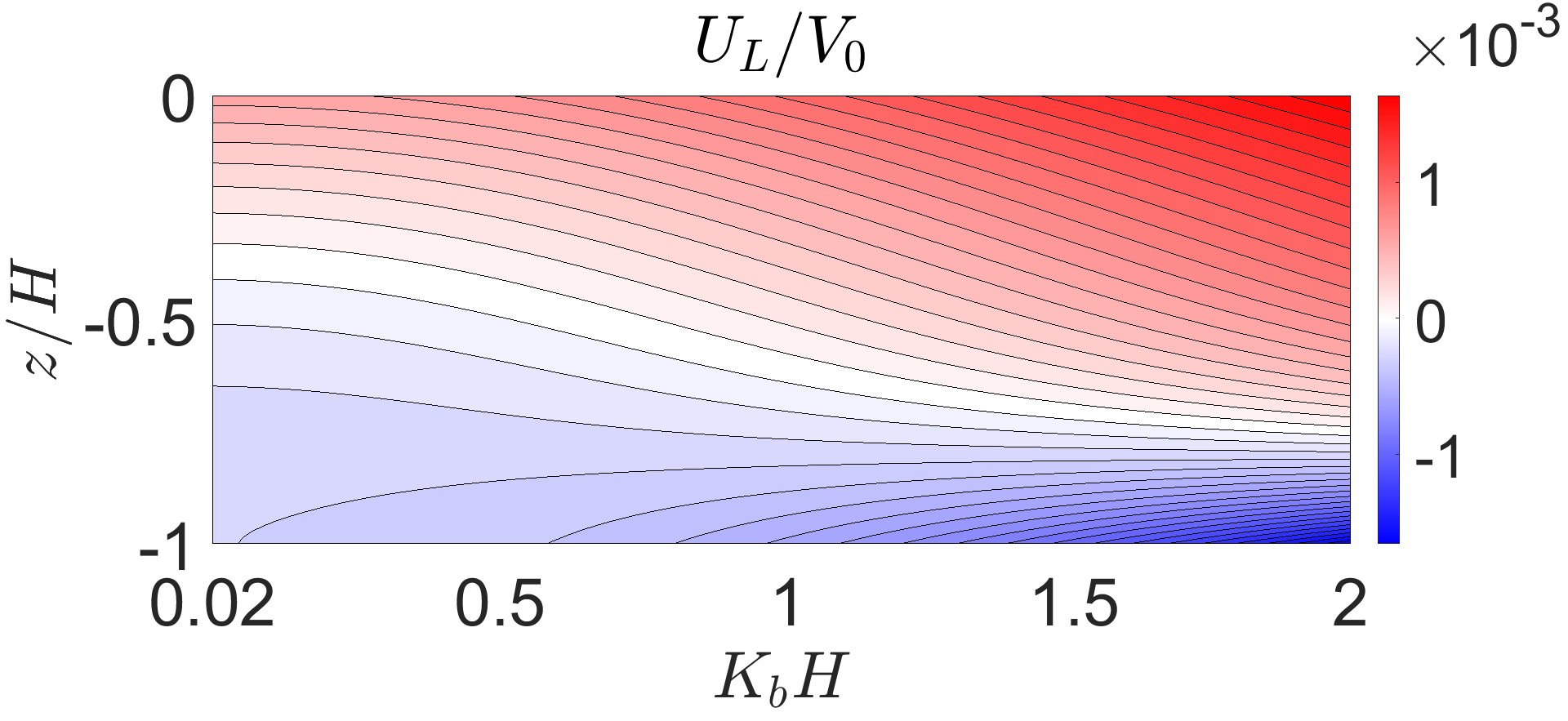}
  \caption{} 
  \end{subfigure}
  \caption{\footnotesize Contour plots in the wavenumber--depth plane of (a)  $\langle u_{SD} \rangle$, (b) $U_{L(SD)}$, (c)  $\langle u_{CBIID} \rangle$, (d) $U_{L(CBIID)}$,  (e) $\langle u_{CD} \rangle$, and (f) $U_{L}$. All velocities have been nondimensionalized by $V_0$. For plots (a,b,e,f): $a=0.01H$, and for plots (e,f): $KH=1$. For plots (c,d,e,f): $\beta=45\degree$, $a_b=0.05H$. For all the plots, $Fr=0.1$ ($V_0>0$). } 
  \label{fig:Lagragian_contours}
\end{figure}

{ We consider realistic nearshore parameters  to calculate  drift and Lagrangian velocities for the three cases. In this regard, observation data from a barred beach near Duck, North Carolina provides the following typical values: $H=2.5$ m and $V_0 =0.5$ m/s \citep{church1993effects,feddersen2000velocity}, leading to $Fr=0.1$.}
$\langle u_{SD} \rangle$ and $U_{L(SD)}$ for case-I are respectively represented by figures \ref{fig:Lagragian_contours}\,({\it a\/}) and \ref{fig:Lagragian_contours}\,({\it b\/}). For these plots, we have chosen a small amplitude ($a/H=0.01$) surface wave whose wavelength ranges between long and intermediate.  Figure \ref{fig:Lagragian_contours}\,({\it b\/}) reveals that $U_{L(SD)}$ is higher for near-surface particles carried by intermediate wavelength surface waves. For long wavelength surface waves, Stokes drift is balanced by the return flow, leading to $U_{L(SD)}\approx 0$.  Case-II shows a very similar variation of CBIID with bottom undulation wavenumber, see figure  \ref{fig:Lagragian_contours}\,({\it c\/}--{\it d\/}).
Nearshore sandbars typically have long ($K_b H \approx 0.07$) to  intermediate ($K_b H\approx 1$)  undulations, and an amplitude of $a_b/H \approx 0.04-0.20$ \citep{dolan1985multiple}, thereby justifying our choice of  $a_b/H=0.05$ and the bottom wavenumber range of $0.02-2$. The crucial difference between figures  \ref{fig:Lagragian_contours}\,({\it b\/}) and \ref{fig:Lagragian_contours}\,({\it d\/}) is that for the latter, high Lagrangian drift (i.e. $U_{L(CBIID)}$) occurs in the neighbourhood of the bottom topography. This is simply because of the fact that   $\langle u_{CBIID} \rangle$ also peaks near the bottom topography, see figure  \ref{fig:parameters_comparison_contours}\,({\it d\/}) or figure  \ref{fig:Lagragian_contours}\,({\it c\/}). Figure \ref{fig:Lagragian_contours}\,({\it e\/}) shows the combined drift
$\langle u_{CD} \rangle =\langle u_{SD} \rangle+ \langle u_{CBIID} \rangle$, and is not limited to $\mathrm{O}(\epsilon)\sim \mathrm{O}(\epsilon_b)$ (as in \S \ref{sec:case3}).
Finally, figure  \ref{fig:Lagragian_contours}\,({\it f\/}) shows the net Lagrangian drift  ($U_L$) for $KH=1$ (i.e. surface wave parameters are held fixed).  If the effect of CBIID was not accounted for, Lagrangian drift would only mean $U_{L(SD)}$, which might lead to erroneous estimates. Figure \ref{fig:Lagragian_contours}\,({\it f\/}) does reveal that the net Lagrangian drift would  vary with $K_bH$  (and also $z$), and the numerical range of $U_{L}$ for a given $K_bH$ could be quite different from that of $U_{L(SD)}$. The difference between  $U_{L(SD)}$ and $U_{L}$ will be more prominent for intermediate wavelength bottom undulations. {The issue is put into perspective in table \ref{table:real_drift_velocities}, where dimensional drift and Lagrangian velocities (in m/s) for particles at the surface, mid-depth, and bottom are provided. Here, both  surface waves and bottom undulations are of intermediate depth, and $\mathrm{O}(\epsilon)\sim \mathrm{O}(\epsilon_b)$. For all depths, $U_{L(CBIID)}$ leads to non-trivial  differences between $U_{L(SD)}$ and $U_{L}$. An interesting feature is observed at the mid-depth, where $U_{L(SD)}$ is negative while  $U_{L}$ is positive. This implies that if CBIID is not taken into consideration, an onshore directed tracer transport could be mistakenly predicted as offshore directed.}

{We note in passing that for situations   where both surface waves and bottom undulations are short (i.e. $KH\gg 1$ and $K_bH \gg 1$), Stokes drift will primarily cause the  transport of floating particles, while CBIID will be instrumental in the transport of heavier particles located near the bottom topography.}

\begin{table}
  \begin{center}
  \def~{\hphantom{0}}
  \color{black}{{\begin{tabular}{lccccccc}
        & \multicolumn{2}{c}{ \textbf{{Case-I}}}   & \multicolumn{2}{c}{\textbf{ {Case-II}}}     & \multicolumn{2}{c}{ \textbf{{Case-III}} }   \\
        $\boldsymbol{z}$ & $\boldsymbol{ \langle u_{SD} \rangle} $  &  $\boldsymbol{U_{L(SD)}}$  & $\boldsymbol{\langle u_{CBIID} \rangle}$   & $\boldsymbol{U_{L(CBIID)}}$   &   $\boldsymbol{\langle u_{CD} \rangle}$   & $\boldsymbol{U_L}$ \\ \hline
        0    & $5.94\times 10^{-4}$  & $3.08\times 10^{-4}$  & $-2.29\times 10^{-4}$ &  $1.83\times 10^{-4}$  & $3.65\times 10^{-4}$   &  $4.91\times 10^{-4}$ \\ 
         $-H/2$    & $2.44\times 10^{-4}$  & $-0.43\times 10^{-4}$  & $-2.90\times 10^{-4}$ &  $1.22\times 10^{-4}$  & $-0.47\times 10^{-4}$   &  $0.80\times 10^{-4}$ \\
        $-H$    & $1.58\times 10^{-4}$  & $-1.28\times 10^{-4}$  & $-5.42\times 10^{-4}$ &  $-1.30\times 10^{-4}$  & $-3.84\times 10^{-4}$   &  $-2.58\times 10^{-4}$ \\ 
\end{tabular}}}
\caption{\footnotesize { Dimensional drift and Lagrangian velocities (in m/s) for the three cases evaluated at $z=0, -H/2, -H$.  Case-I:  $KH=1$ and $a=0.01H$ (with flat bottom topography), case-II:  $K_b H=1$, $a_b=0.05H$, and $\beta=45\degree$ (with no surface waves), case-III:  $KH=1$, $a=0.01H$, $K_b H=1$, $a_b=0.05H$, and $\beta=45\degree$ (previous two cases combined). For all cases,  $V_0=0.5$ m/s and $H=2.5$ m, leading to $Fr=0.1$.}}
  \label{table:real_drift_velocities}
  \end{center}
\end{table}

\vspace{-3mm}
{\section{Summary and conclusions}
\label{sec:conclusion}}                                   

In this paper we have shown that in free surface flows,  cross-stream drift velocity can be generated via the interactions between a uniform, streamwise current and small amplitude {wavy} bottom topography. This phenomenon is especially expected to have non-trivial implications in shallow free surface flows with strong currents (i.e. $Fr=\mathrm{O}(1)$). 
{Focusing on a simple model of the nearshore environment, we show that the proposed drift mechanism (referred to as `CBIID') might  play an important role in the cross-shelf transport of nearshore tracers  like planktons, pollutants, pathogens, and sediments.} In  coastal environments, cross-shelf tracer transport is often mediated through the Stokes drift mechanism -- mass transport by surface waves. Mathematically, Stokes drift results from the transient, homogeneous part of the governing system of equations. Au contraire, the steady, non-homogeneous  part of the governing equations give rise to CBIID. {It is specifically shown to be an important contributor to the net Lagrangian drift in nearshore environments having oblique sandbars of intermediate wavelengths. }

Depending on the angle between the alongshore current and the bottom {topography} wave-vector, CBIID can lead to  onshoreward or offshoreward tracer transport. The CBIID velocity in the cross-shelf direction, $\langle \boldsymbol{u}_{CBIID} \rangle$, is maximum when the bottom {topography} wave-vector approximately makes an angle of $\pi/4$ with the shoreline. $\langle \boldsymbol{u}_{CBIID} \rangle$ also increases with particle's initial depth (hence submerged particles will experience it more strongly than floating particles), the magnitudes of the alongshore current, and bottom  {topography's} amplitude. {Moreover, $\langle \boldsymbol{u}_{CBIID} \rangle$ near the free surface increases with an increase in the bottom undulation's wavelength, while near the bottom, $\langle \boldsymbol{u}_{CBIID} \rangle$ decreases with an increase in bottom undulation's wavelength. Hence in situations where both surface waves and bottom topography have short wavelengths, Stokes drift is expected to cause  transport of floating particles, while CBIID will transport heavier particles located near the bottom topography.}

{For the situations we investigated, particle motions due to surface waves or wavy bathymetry occurred in the presence of a background flow (for wavy bathymetry, background flow is mandatory), as expected in a realistic environment. A minor but essential outcome of this work is that small-excursion approximation, used for solving  pathline equations, is \emph{only} valid when the analysis is performed in a reference frame moving with the background current. Hence, in order to apply the well-known expression for Stokes drift (obtained in a rest frame) in a realistic oceanic scenario,  proper variable transformations are necessary.}
We also investigated whether  \emph{a-priori} assumption of small-excursion approximation, which is essential for providing a highly accurate expression for Stokes drift, also provides  highly accurate estimates of CBIID. We show that for moderate and long wavelength bottom {topography}, particle excursions may not be small, hence {the} small-excursion approximation does not provide accurate estimates. To circumvent this issue, we introduced the `$z$-bounded approximation', which (like {the} small-excursion approximation) provides simple expressions for the drift velocity and time period. {Since the $z$-bounded approximation does not demand small-excursions in the horizontal plane, the analysis can be applied in a rest-frame.}
The $z$-bounded approximation is found to be highly accurate and indistinguishable from the exact solution in general.
However, since $z$-excursions increase with water depth, the $z$-bounded approximation can show differences with the exact solution in the vicinity of a bottom topography whose amplitude is not too small { (e.g. $a_b/H \gtrsim 0.1$).}

We have limited our study to a model nearshore environment consisting of uniform alongshore current, onshore propagating surface waves, and monochromatic   {wavy} bottom making an oblique angle with the shoreline. {Extension to polychromatic bathymetry, so as to include realistic complex seabed undulations, would be a straightforward extension of our analysis (addition of various Fourier components)}. While we have included Eulerian return flow in a simplified sense,   many important nearshore processes, for example,  boundary layers, wind forcing, wave breaking, and transient rip currents have been ignored. {Moreover in practical  scenarios, both $V_0$ and $H$ would vary in the cross-shelf direction. Our simplified model would produce reasonably accurate results for scenarios where the variations in $V_0$ and $H$ are `mild', i.e. the horizontal stretch can accommodate multiple bottom undulations without much changes in $V_0$ and $H$.  }

The key advantage of our simplified analysis lies in underpinning the drift mechanism arising through current-bathymetry interactions, which might not be evident if various complex processes are taken into consideration. Moreover, the model environment considered in our study represents a fairly generic free surface flow scenario, and hence can be extended to other natural water bodies like rivers and estuaries. Realistic parametric analysis of the nearshore environment has revealed that in situations where the Stokes drift  and CBIID velocities have comparable magnitudes, they are order separated both in length and time scales. In this scenario, the Lagrangian drift can be  represented as the sum of {the net} Eulerian return flow, Stokes drift, and CBIID. 
Furthermore, we show that if {topographic effects} are not taken into consideration (i.e. CBIID is absent), the prediction of the net Lagrangian drift might be erroneous. Hence, obtaining high resolution coastal bathymetry map is essential for accurate prediction of nearshore tracer transport.
While bathymetry reconstruction is an ongoing challenge, recent efforts have provided bathymetry maps at $15$ arc-seconds resolution  \citep{tozer2019global}. In addition, the upcoming Surface Water and Ocean Topography (SWOT) satellite mission is expected to reconstruct bathymetry at unprecedented resolution.

\vspace{2mm}

\section*{Acknowledgment}
A. Gupta thanks Commonwealth Split-site Ph.D. scholarship for funding support. The authors would like to thank the Associate Editor Prof. James Kirby and the anonymous reviewers for their constructive suggestions and comments, which have led to significant improvements in the manuscript.

\section*{Declaration of interests}
The authors report no conflict of interest. 
\vspace{-5mm}
\appendix


\section{Evaluation of combined drift in case-III using the small-excursion approximation:}
\label{appendixsec:small_excursion} 

The \emph{approximate}-CD velocities in the $x$-, $y$-, and $z$- directions are respectively derived from \eqref{eq:combined_drift_formula}, which are as follows

\begin{subequations}
\begin{align}
\langle u_{aCD} \rangle &= \underbrace{\langle \widetilde{X}_u u_{u,X}^{(1)}+ \widetilde{Y}_u u_{u,Y}^{(1)}+ \widetilde{Z}_u u_{u,Z}^{(1)}\rangle}_{\text{\clap{Term-1}}} + \underbrace{\langle \widetilde{X}_s u_{s,X}^{(1)}+ \widetilde{Y}_s u_{s,Y}^{(1)}+ \widetilde{Z}_s u_{s,Z}^{(1)}\rangle}_{\text{\clap{Term-2}}} +  \nonumber \\& \underbrace{ \langle \widetilde{X}_u u_{s,X}^{(1)}+ \widetilde{Y}_u u_{s,Y}^{(1)}+ \widetilde{Z}_u u_{s,Z}^{(1)}\rangle}_{\text{\clap{Term-3}}} + \underbrace{\langle \widetilde{X}_s u_{u,X}^{(1)}+ \widetilde{Y}_s u_{u,Y}^{(1)}+ \widetilde{Z}_s u_{u,Z}^{(1)} \rangle}_{\text{\clap{Term-4}}}, \label{app eq:u_aCD}\\
\langle v_{aCD} \rangle &= \underbrace{\langle \widetilde{X}_u v_{u,X}^{(1)}+ \widetilde{Y}_u v_{u,Y}^{(1)}+ \widetilde{Z}_u v_{u,Z}^{(1)}\rangle}_{\text{\clap{Term-5}}} + \underbrace{\langle \widetilde{X}_s v_{s,X}^{(1)}+ \widetilde{Y}_s v_{s,Y}^{(1)}+ \widetilde{Z}_s v_{s,Z}^{(1)}\rangle}_{\text{\clap{Term-6}}} +  \nonumber \\& \underbrace{\langle \widetilde{X}_u v_{s,X}^{(1)}+ \widetilde{Y}_u v_{s,Y}^{(1)}+ \widetilde{Z}_u v_{s,Z}^{(1)}\rangle}_{\text{\clap{Term-7}}} + \underbrace{\langle \widetilde{X}_s v_{u,X}^{(1)}+ \widetilde{Y}_s v_{u,Y}^{(1)}+ \widetilde{Z}_s v_{u,Z}^{(1)} \rangle}_{\text{\clap{Term-8}}}, \label{app eq:v_aCD}\\
\langle w_{aCD} \rangle &= \underbrace{\langle \widetilde{X}_u w_{u,X}^{(1)}+ \widetilde{Y}_u w_{u,Y}^{(1)}+ \widetilde{Z}_u w_{u,Z}^{(1)}\rangle}_{\text{\clap{Term-9}}} + \underbrace{\langle \widetilde{X}_s w_{s,X}^{(1)}+ \widetilde{Y}_s w_{s,Y}^{(1)}+ \widetilde{Z}_s w_{s,Z}^{(1)}\rangle}_{\text{\clap{Term-10}}} +  \nonumber \\& \underbrace{ \langle \widetilde{X}_u w_{s,X}^{(1)}+ \widetilde{Y}_u w_{s,Y}^{(1)}+ \widetilde{Z}_u w_{s,Z}^{(1)}\rangle}_{\text{\clap{Term-11}}} + \underbrace{\langle \widetilde{X}_s w_{u,X}^{(1)}+ \widetilde{Y}_s w_{u,Y}^{(1)}+ \widetilde{Z}_s w_{u,Z}^{(1)} \rangle}_{\text{\clap{Term-12}}}. \label{app eq:w_aCD}
\end{align}
\end{subequations}
where $\langle \ldots \rangle$ denotes averaging over one time period in the moving frame, $T_{aCD}$. We have defined $T_{aCD}$ such that
$T_{aCD} \approx n T_{aCBIID} \approx m \overline{T}$, where $m,n \in \mathbb{Z}^+$. The various terms appearing in \eqref{app eq:u_aCD}--\eqref{app eq:w_aCD} are as follows:
\begin{equation*}
    \text{Term-1}=\langle u_{SD} \rangle, \,\,\text{Term-5}=\langle v_{SD} \rangle,\,\, \text{Term-9}=\langle w_{SD} \rangle,
\end{equation*}
where $u_{SD},\,v_{SD},$ and $w_{SD}$ are respectively given in \eqref{eq:u_sd}, \eqref{eq:v_sd}, and \eqref{eq:w_sd}. Likewise,
\begin{equation*}
    \text{Term-2}=\langle u_{aCBIID} \rangle, \,\,\text{Term-6}=\langle v_{aCBIID} \rangle,\,\, \text{Term-10}=\langle w_{aCBIID} \rangle,
\end{equation*}
where $u_{aCBIID},\,v_{aCBIID},$ and $w_{aCBIID}$ are respectively given in \eqref{eq:u_aCBIID}, \eqref{eq:v_aCBIID}, and \eqref{eq:w_aCBIID}. Additionally, 
\begin{align*}
    \text{Term-3}+\text{Term-4}&= \frac{a (k_b V_0 l_b-k \overline{\omega})}{K V_0 l_b \sinh (K H)}  \{ (k k_b+ll_b) {\mathbb{P}}\cosh K(z_0+H)+ K K_b {\mathbb{Q}}\sinh K(z_0+H) \} \mathcal{I},\\
    \text{Term-7}+\text{Term-8}&= \frac{a (l_b V_0 l_b-l \overline{\omega})}{K V_0 l_b \sinh (K H)}  \{ (k k_b+ll_b) {\mathbb{P}}\cosh K(z_0+H)+ K K_b {\mathbb{Q}}\sinh K(z_0+H) \} \mathcal{I},\\
    \text{Term-11}+\text{Term-12}&= \frac{a }{V_0 l_b \sinh (K H)} \big[ \{\overline{\omega}(k k_b+ll_b)+V_0 l_b K K_b\} {\mathbb{P}}  (\mathcal{P}-\mathcal{R}) \sinh K(z_0+H) + \\ & \quad K K_b (\overline{\omega}+V_0 l_b) {\mathbb{Q}}  (\mathcal{P}+\mathcal{R})\cosh K(z_0+H) \big],
\end{align*}
where, 
\begin{align*}
\mathcal{I}= \frac{\cos\{(k-k_b)x_0+(l-l_b) y_0- \pi(m+n)\} \sin\{\pi(m+n)\}}{2 \pi (m+n)} + \\ \frac{\cos\{(k+k_b)x_0+(l+l_b) y_0- \pi(m-n)\} \sin\{\pi(m-n)\}}{2 \pi (m-n)}, \\
\mathcal{P}= \frac{\sin\{(k+k_b)x_0+(l+l_b) y_0- \pi(m-n)\} \sin\{\pi(m-n)\}}{2 \pi (m-n)}, \\
\mathcal{R}= \frac{\sin\{(k-k_b)x_0+(l-l_b) y_0- \pi(m+n)\} \sin\{\pi(m+n)\}}{2 \pi (m+n)}. 
\end{align*} 
Since $m,n \in \mathbb{Z}^+$, we have $\mathcal{I}=\mathcal{P}=\mathcal{R}=0$, leading to zero values for Terms -3, -4, -7, -8, -11, and -12. This finally yields
\begin{equation}
\boldsymbol{\langle u_{aCD} \rangle} = \boldsymbol{\langle u_{SD} \rangle}+\boldsymbol{\langle u_{aCBIID} \rangle}.
\label{app eq:app_CD}
\end{equation}

\bibliographystyle{jfm}
\bibliography{jfm-instructions}

\begin{thebibliography}{35}
\expandafter\ifx\csname natexlab\endcsname\relax\def\natexlab#1{#1}\fi
\def\au#1{#1} \def\ed#1{#1} \def\yr#1{#1}\def\at#1{#1}\def\jt#1{\textit{#1}}
  \def\bt#1{#1}\def\bvol#1{\textbf{#1}} \def\vol#1{#1} \def\pg#1{#1}
  \def\publ#1{#1}\def\arxiv#1{#1}\def\org#1{#1}\def\st#1{\textit{#1}}

\bibitem[Akselsen \& Ellingsen(2019)]{akselsen2019sheared}
{\sc \au{Akselsen, A.~H.} \& \au{Ellingsen, S.~{\AA}.}} \yr{2019}  \at{Sheared
  free-surface flow over three-dimensional obstructions of finite amplitude}.
  \jt{J. Fluid Mech.}  \bvol{878},  \pg{740--767}.

\bibitem[Bakhtyar {\em et~al.\/}(2016)Bakhtyar, Dastgheib, Roelvink \&
  Barry]{bakhtyar2016impacts}
{\sc \au{Bakhtyar, R.}, \au{Dastgheib, A.}, \au{Roelvink, D.} \& \au{Barry,
  D.~A.}} \yr{2016}  \at{Impacts of wave and tidal forcing on 3{D} nearshore
  processes on natural beaches. {P}art {I}: {F}low and turbulence fields}.
  \jt{Ocean Syst. Eng.}  \bvol{6},  \pg{23--60}.

\bibitem[Van~den Bremer \& Breivik(2017)]{van2017stokes}
{\sc \au{Van~den Bremer, T.~S.} \& \au{Breivik, {\O}.}} \yr{2017}  \at{Stokes
  drift}.  \jt{Philos. Trans. R. Soc. A}  \bvol{376}~(2111),  \pg{20170104}.

\bibitem[Brink(2016)]{brink2016cross}
{\sc \au{Brink, K.~H.}} \yr{2016}  \at{Cross-shelf exchange}.  \jt{Annu. Rev.
  Mar. Sci.}  \bvol{8},  \pg{59--78}.

\bibitem[Brown {\em et~al.\/}(2015)Brown, MacMahan, Reniers \&
  Thornton]{brown2015field}
{\sc \au{Brown, J.~A.}, \au{MacMahan, J.~H.}, \au{Reniers, A.~J.} \&
  \au{Thornton, E.~B.}} \yr{2015}  \at{Field observations of surf zone--inner
  shelf exchange on a rip-channeled beach}.  \jt{J. Phys. Oceanogr.}
  \bvol{45}~(9),  \pg{2339--2355}.

\bibitem[Church \& Thornton(1993)]{church1993effects}
{\sc \au{Church, J.~C.} \& \au{Thornton, E.~B.}} \yr{1993}  \at{Effects of
  breaking wave induced turbulence within a longshore current model}.
  \jt{Coast. Eng.}  \bvol{20}~(1-2),  \pg{1--28}.

\bibitem[Clamond(2007)]{clamond2007lagrangian}
{\sc \au{Clamond, D.}} \yr{2007}  \at{On the {L}agrangian description of steady
  surface gravity waves}.  \jt{J. Fluid Mech.}  \bvol{589},  \pg{433--454}.

\bibitem[Constantin {\em et~al.\/}(2008)Constantin, Ehrnstr{\"o}m \&
  Villari]{constantin2008particle2}
{\sc \au{Constantin, A.}, \au{Ehrnstr{\"o}m, M.} \& \au{Villari, G.}} \yr{2008}
   \at{Particle trajectories in linear deep-water waves}.  \jt{Nonlinear Anal.
  Real World Appl.}  \bvol{9}~(4),  \pg{1336--1344}.

\bibitem[Constantin \& Villari(2008)]{constantin2008particle}
{\sc \au{Constantin, A.} \& \au{Villari, G.}} \yr{2008}  \at{Particle
  trajectories in linear water waves}.  \jt{J. Math. Fluid Mech.}
  \bvol{10}~(1),  \pg{1--18}.

\bibitem[Dolan \& Dean(1985)]{dolan1985multiple}
{\sc \au{Dolan, T.~J.} \& \au{Dean, R.~G.}} \yr{1985}  \at{Multiple longshore
  sand bars in the upper {C}hesapeake {B}ay}.  \jt{Estuar. Coast. Shelf Sci.}
  \bvol{21}~(5),  \pg{727--743}.

\bibitem[Dommermuth \& Yue(1987)]{dommermuth1987high}
{\sc \au{Dommermuth, D.~G.} \& \au{Yue, D. K.~P.}} \yr{1987}  \at{A high-order
  spectral method for the study of nonlinear gravity waves}.  \jt{J. Fluid
  Mech.}  \bvol{184},  \pg{267--288}.

\bibitem[Fan {\em et~al.\/}(2021)Fan, Zheng, Tao \& Liu]{fan2021upstream}
{\sc \au{Fan, J.}, \au{Zheng, J.}, \au{Tao, A.} \& \au{Liu, Y.}} \yr{2021}
  \at{Upstream-propagating waves induced by steady current over a rippled
  bottom: theory and experimental observation}.  \jt{J. Fluid Mech.}
  \bvol{910}.

\bibitem[Feddersen {\em et~al.\/}(2000)Feddersen, Guza, Elgar \&
  Herbers]{feddersen2000velocity}
{\sc \au{Feddersen, F.}, \au{Guza, R.~T.}, \au{Elgar, S.} \& \au{Herbers, T.
  H.~C.}} \yr{2000}  \at{Velocity moments in alongshore bottom stress
  parameterizations}.  \jt{J. Geophys. Res. Oceans}  \bvol{105}~(C4),
  \pg{8673--8686}.

\bibitem[Gelfenbaum(2005)]{gelfenbaum2005coastal}
{\sc \au{Gelfenbaum, G.}} \yr{2005}  \at{Coastal currents}.  \jt{Encyclopedia
  of Coastal Science, Springer}  \pg{pp. 259--260}.

\bibitem[Gupta \& Guha(2021)]{gupta2021modified}
{\sc \au{Gupta, A.} \& \au{Guha, A.}} \yr{2021}  \at{Modified {S}tokes drift
  due to resonant interactions between surface waves and corrugated sea floor
  with and without a mean current}.  \jt{Phys. Rev. Fluids}  \bvol{6}~(2),
  \pg{024801}.

\bibitem[Henry(2007)]{henry2007particle}
{\sc \au{Henry, D.}} \yr{2007}  \at{Particle trajectories in linear periodic
  capillary and capillary--gravity water waves}.  \jt{Philos. Trans. Royal Soc.
  A}  \bvol{365}~(1858),  \pg{2241--2251}.

\bibitem[Kennedy(1963)]{kennedy1963mechanics}
{\sc \au{Kennedy, J.~F.}} \yr{1963}  \at{The mechanics of dunes and antidunes
  in erodible-bed channels}.  \jt{J. Fluid Mech.}  \bvol{16}~(4),
  \pg{521--544}.

\bibitem[Kirby(1988)]{kirby1988current}
{\sc \au{Kirby, J.~T.}} \yr{1988}  \at{Current effects on resonant reflection
  of surface water waves by sand bars}.  \jt{J. Fluid Mech.}  \bvol{186},
  \pg{501--520}.

\bibitem[Kumar \& Feddersen(2017)]{kumar2017effect}
{\sc \au{Kumar, N.} \& \au{Feddersen, F.}} \yr{2017}  \at{The effect of
  {Stokes} drift and transient rip currents on the inner shelf. {P}art {I}: No
  stratification}.  \jt{J. Phys. Oceanogr.}  \bvol{47}~(1),  \pg{227--241}.

\bibitem[Kundu {\em et~al.\/}(2016)Kundu, Cohen \& Dowling]{kundu20fluid}
{\sc \au{Kundu, P.~K.}, \au{Cohen, I.~M.} \& \au{Dowling, D.~R.}} \yr{2016}
  {\em Fluid Mechanics (Sixth Edition)\/}.  \publ{Academic Press}.

\bibitem[Lamb(1932)]{lamb1932}
{\sc \au{Lamb, H.}} \yr{1932} {\em Hydrodynamics\/}.  \publ{University Press}.

\bibitem[Lentz {\em et~al.\/}(2008)Lentz, Fewings, Howd, Fredericks \&
  Hathaway]{lentz2008observations}
{\sc \au{Lentz, S.~J.}, \au{Fewings, M.}, \au{Howd, P.}, \au{Fredericks, J.} \&
  \au{Hathaway, K.}} \yr{2008}  \at{Observations and a model of undertow over
  the inner continental shelf}.  \jt{J. Phys. Oceanogr.}  \bvol{38}~(11),
  \pg{2341--2357}.

\bibitem[Lentz \& Fewings(2012)]{lentz2012wind}
{\sc \au{Lentz, S.~J.} \& \au{Fewings, M.~R.}} \yr{2012}  \at{The wind-and
  wave-driven inner-shelf circulation}.  \jt{Annu. Rev. Mar. Sci.}  \bvol{4},
  \pg{317--343}.

\bibitem[Longuet-Higgins(1953)]{longuet1953mass}
{\sc \au{Longuet-Higgins, M.~S.}} \yr{1953}  \at{Mass transport in water
  waves}.  \jt{Philos. Trans. Royal Soc. A}  \bvol{245}~(903),  \pg{535--581}.

\bibitem[Longuet-Higgins(1987)]{longuet1987lagrangian}
{\sc \au{Longuet-Higgins, M.~S.}} \yr{1987}  \at{Lagrangian moments and mass
  transport in {S}tokes waves}.  \jt{J. Fluid Mech.}  \bvol{179},
  \pg{547--555}.

\bibitem[O'Dea {\em et~al.\/}(2021)O'Dea, Kumar \& Haller]{o2021simulations}
{\sc \au{O'Dea, A.}, \au{Kumar, N.} \& \au{Haller, M.~C.}} \yr{2021}
  \at{Simulations of the surf zone eddy field and cross-shore exchange on a
  nonidealized bathymetry}.  \jt{J. Geophys. Res. Oceans}  \bvol{126}~(5).

\bibitem[Peregrine(1976)]{peregrine1976interaction}
{\sc \au{Peregrine, D.~H.}} \yr{1976}  \at{Interaction of water waves and
  currents}.  \jt{Adv. Appl. Mech.}  \bvol{16},  \pg{9--117}.

\bibitem[Raj \& Guha(2019)]{raj_guha_2019}
{\sc \au{Raj, R.} \& \au{Guha, A.}} \yr{2019}  \at{On {B}ragg resonances and
  wave triad interactions in two-layered shear flows}.  \jt{J. Fluid Mech.}
  \bvol{867},  \pg{482–515}.

\bibitem[Rao(2004)]{rao2004flow}
{\sc \au{Rao, T. V.~N.}} \yr{2004}  \at{Flow field in the inner shelf along the
  {C}entral {E}ast {C}oast of {I}ndia during the southwest monsoon season}.
  \jt{J. Coast. Res.}  \bvol{20}~(3),  \pg{814--827}.

\bibitem[Ribas {\em et~al.\/}(2003)Ribas, Falqu{\'e}s \&
  Montoto]{ribas2003nearshore}
{\sc \au{Ribas, F.}, \au{Falqu{\'e}s, A.} \& \au{Montoto, A.}} \yr{2003}
  \at{Nearshore oblique sand bars}.  \jt{J. Geophys. Res. Oceans}
  \bvol{108}~(C4).

\bibitem[Sammarco {\em et~al.\/}(1994)Sammarco, Mei \&
  Trulsen]{sammarco1994nonlinear}
{\sc \au{Sammarco, P.}, \au{Mei, C.~C.} \& \au{Trulsen, K.}} \yr{1994}
  \at{Nonlinear resonance of free surface waves in a current over a sinusoidal
  bottom: a numerical study}.  \jt{J. Fluid Mech.}  \bvol{279},  \pg{377--405}.

\bibitem[Stokes(1847)]{stokes1847}
{\sc \au{Stokes, G.~G.}} \yr{1847}  \at{On the theory of oscillatory waves}.
  \jt{Trans. Camb. Philos. Soc.}  \bvol{8},  \pg{441}.

\bibitem[Thomson(1886)]{thomson1886lx}
{\sc \au{Thomson, W. (Lord~Kelvin)}} \yr{1886}  \at{On stationary waves in
  flowing water}.  \jt{Edinb. Dubl. Phil. Mag. J. Sci.}  \bvol{22-23},  \pg{t.
  xxii, pp. 353, 445, 517}.

\bibitem[Tozer {\em et~al.\/}(2019)Tozer, Sandwell, Smith, Olson, Beale \&
  Wessel]{tozer2019global}
{\sc \au{Tozer, B.}, \au{Sandwell, D.~T.}, \au{Smith, W.H.F.}, \au{Olson, C},
  \au{Beale, J.~R.} \& \au{Wessel, P}} \yr{2019}  \at{Global bathymetry and
  topography at 15 arc sec: {SRTM}15+}.  \jt{Earth Space Sci.}  \bvol{6}~(10),
  \pg{1847--1864}.

\bibitem[Ursell(1953)]{ursell1953long}
{\sc \au{Ursell, F.}} \yr{1953}  \at{The long-wave paradox in the theory of
  gravity waves}.  \jt{Math. Proc. Camb. Philos. Soc}  \bvol{49}~(4),
  \pg{685--694}.

\end{thebibliography}


\end{document}